\documentclass[conference, 10pt, letterpaper]{IEEEtran}

\usepackage{tikz}
\usepackage{pgfplots}
\pgfplotsset{compat=1.18}
\usepackage{amsmath,amssymb,amsfonts}
\usepackage[cmintegrals]{newtxmath}

\usepackage{interval}
\usepackage{algorithmic}
\usepackage{algorithm}
\usepackage{array}
\usepackage[caption=false,font=normalsize,labelfont=sf,textfont=sf]{subfig}
\usepackage{cite}
\usepackage{graphicx,color}
\graphicspath{{./img/}}
\usepackage{wrapfig}
\usepackage{textcomp}
\usepackage{stfloats}
\usepackage[hyphens]{url}
\usepackage[hidelinks,breaklinks=true]{hyperref}
\usepackage[capitalise]{cleveref}
\usepackage{bmpsize}
\usepackage[nohyperlinks,nolist]{acronym} 
\usepackage{tabularx}
\newcolumntype{Y}{>{\centering\arraybackslash}X}
\newcolumntype{L}{>{\arraybackslash}X}
\newcolumntype{R}{>{\raggedleft\arraybackslash}X}
\newcolumntype{C}[1]{>{\centering\arraybackslash}p{#1}}
\usepackage{booktabs}
\usepackage{multirow}
\usepackage[per-mode=symbol,group-separator={,},group-minimum-digits={3}]{siunitx}
\usepackage{listings}
\usepackage{enumitem}

\usepackage{pifont}

\usepackage{xspace}
\usepackage{xstring}
\newcommand{\etal}{\textit{et~al.}~}
\newcommand{\eg}{\textit{e.g.,}~}
\newcommand{\ie}{\textit{i.e.,}~}
\newcommand{\cf}{\textit{cf.}~}

\newcommand{\one}{{\em(i)}~}
\newcommand{\two}{{\em(ii)}~}
\newcommand{\three}{{\em(iii)}~}
\newcommand{\four}{{\em(iv)}~}

\usepackage{textcmds}

\definecolor{CoreGray}{HTML}{BFBFBF}
\definecolor{CoreBlack}{HTML}{333333}
\definecolor{CoreDarkGray}{HTML}{5F5F5F}
\definecolor{CoreBlue}{HTML}{002E7D}
\definecolor{CoreGreen}{HTML}{008000}
\definecolor{CoreGreen2}{HTML}{6AAC8E}
\definecolor{CoreRed}{HTML}{C80000}
\definecolor{CoreYellow}{HTML}{E6AC00}
\definecolor{CoreWhite}{HTML}{FFFFFF}
\definecolor{CoreMagenta}{HTML}{7030A0}
\colorlet{LightCoreGray}{CoreGray!30}
\colorlet{LightCoreBlack}{CoreBlack!20}
\colorlet{LightCoreBlue}{CoreBlue!20}
\colorlet{LightCoreGreen}{CoreGreen!30}
\colorlet{LightCoreRed}{CoreRed!20}
\colorlet{LightCoreYellow}{CoreYellow!20}
\colorlet{LightCoreWhite}{CoreWhite!20}

\newcommand\codeword[1]{\mbox{\textbf{\texttt{\textcolor{CoreBlack}{#1}}}}}
\usepackage{mathtools}

\newcommand{\customparagraph}[1]{\vspace{3pt}\noindent\textbf{#1}\IfEndWith{#1}{.}{\hspace{0.25em}}{}}

\begin{document}

\date{}

\title{Building Automotive Security on Internet Standards: \\In-Car Service Authentication with DNSSEC, DANE, and DANCE}

\DeclareRobustCommand*{\fixautherrefmark}[1]{%
  \raisebox{0pt}[0pt][0pt]{\textnormal{{\textsuperscript{\scriptsize #1}}}}%
}
\author{\IEEEauthorblockN{%
  Timo~Salomon\fixautherrefmark{1,2}, 
  Mehmet~Mueller\fixautherrefmark{2},
  Philipp~Meyer\fixautherrefmark{2},
  and~Thomas~C.~Schmidt\fixautherrefmark{2}}
\IEEEauthorblockA{\fixautherrefmark{1}
{Faculty of Computer Science, Dresden University of Technology, 01187 Dresden, Germany}}
\IEEEauthorblockA{\fixautherrefmark{2}
{Department of Computer Science, Hamburg University of Applied Sciences, 20099 Hamburg, Germany}}
Corresponding Author: Timo Salomon (e-mail: \href{mailto:timo.salomon@haw-hamburg.de}{timo.salomon@haw-hamburg.de}).
}

\maketitle

\begin{abstract}

The automotive industry trends to softwarizing cars according to flexible, service-oriented architectures. 
At the same time, connected cars become networked computer systems but lack proper security measures. 
In this paper, we propose to authenticate and authorize in-vehicle services by integrating DNSSEC, DANE, and DANCE with automotive middleware. Our approach decouples the cryptographic authentication of the service from that of the service deployment with the help of DNSSEC and thereby largely simplifies key management. 
We propose to authenticate in-vehicle services by certificates that are solely generated by the service suppliers but published on deployment via DNSSEC TLSA records solely signed by the OEM. 
We further show how these established Internet standards secure the service discovery and invocation, ensure interoperability with various protocols, and scale management of credentials for millions of connected vehicles at robust security levels. 
We back our design proposal by a security analysis using the STRIDE threat model and by evaluations in an emulation of a realistic in-vehicle setup that demonstrate its effectiveness. 
\end{abstract}

\begin{IEEEkeywords}
Automotive Security, DNSSEC, Certificate Management, Authentication.
\end{IEEEkeywords}

\begin{acronym}
	\acro{ACC}[ACC]{Adaptive Cruise Control}
	\acro{ACDC}[ACDC]{Automotive Cyber Defense Center}
	\acro{ACL}[ACL]{Access Control List}
	\acro{AD}[AD]{Anomaly Detection}
	\acro{ADAS}[ADAS]{Advanced Driver Assistance System}
	\acro{ADS}[ADS]{Anomaly Detection System}
	\acroplural{ADS}[ADSs]{Anomaly Detection Systems}
	\acro{ALPN}[ALPN]{Application Layer Protocol Number}
	\acro{API}[API]{Application Programming Interface}
	\acro{AUTOSAR}[AUTOSAR]{AUTomotive Open System ARchitecture}
	\acro{AVB}[AVB]{Audio Video Bridging}
	\acro{ARP}[ARP]{Address Resolution Protocol}
	\acro{ATS}[ATS]{Asynchronous Traffic Shaping}
	\acro{BE}[BE]{Best-Effort}
	\acro{C2X}[Car2X]{Car-to-Everything}
	\acro{CA}[CA]{Certification Authority}
	\acroplural{CA}[CAs]{Certification Authorities}
	\acro{CAN}[CAN]{Controller Area Network}
	\acro{CBM}[CBM]{Credit Based Metering}
	\acro{CBS}[CBS]{Credit Based Shaper}
	\acro{CNC}[CNC]{Central Network Controller}
	\acro{CUC}[CUC]{Central User Configuration}
	\acro{CMI}[CMI]{Class Measurement Interval}
	\acro{CoRE}[CoRE]{Communication over Realtime Ethernet}
	\acro{CT}[CT]{Cross Traffic}
	\acro{CM}[CM]{Communication Matrix}
	\acro{DANE}[DANE]{DNS-Based Authentication of Named Entities}
	\acro{DANCE}[DANCE]{DANE Authentication for Network Clients Everywhere}
	\acro{DoS}[DoS]{Denial of Service}
	\acro{DDoS}[DDoS]{Distributed \acl{DoS}}
	\acro{DDS}[DDS]{Data Distribution Service}
	\acro{DH}[DH]{Diffie-Hellman}
	\acro{DNS}[DNS]{Domain Name System}
	\acro{DNSSEC}[DNSSEC]{Domain Name System Security Extensions}
	\acro{DPI}[DPI]{Deep Packet Inspection}
	\acro{DNC}[DNC]{Deterministic Network Calculus}
	\acro{DTLS}[DTLS]{Datagram Transport Layer Security}
	\acro{E/E}[E/E]{Electrical/Electronic}
	\acro{ECU}[ECU]{Electronic Control Unit}
	\acroplural{ECU}[ECUs]{Electronic Control Units}
	\acro{FN}[FN]{False Negative}
	\acro{FP}[FP]{False Positive}
	\acro{FDTI}[FDTI]{Fault Detection Time Interval}
	\acro{FHTI}[FHTI]{Fault Handling Time Interval}
	\acro{FRTI}[FRTI]{Fault Reaction Time Interval}
	\acro{FTTI}[FTTI]{Fault Tolerant Time Interval}
	\acro{GCL}[GCL]{Gate Control List}
	\acro{HTTP}[HTTP]{Hypertext Transfer Protocol}
	\acro{HMI}[HMI]{Human-Machine Interface}
	\acro{HPC}[HPC]{High-Performance Controller}
	\acro{HSM}[HSM]{Hardware Security Module}
	\acro{IA}[IA]{Industrial Automation}
	\acro{IAM}[IAM]{Identity- and Access Management}
	\acro{ICT}[ICT]{Information and Communication Technology}
	\acro{IDS}[IDS]{Intrusion Detection System}
	\acroplural{IDS}[IDSs]{Intrusion Detection Systems}
	\acro{IEEE}[IEEE]{Institute of Electrical and Electronics Engineers}
	\acro{IGMP}[IGMP]{Internet Group Management Protocol}
	\acro{IoT}[IoT]{Internet of Things}
	\acro{IP}[IP]{Internet Protocol}
	\acro{IVN}[IVN]{In-Vehicle Network}
	\acroplural{IVN}[IVNs]{In-Vehicle Networks}
	\acro{KSK}[KSK]{Key Signing Key}
	\acro{LIN}[LIN]{Local Interconnect Network}
	\acro{MAC}[MAC]{Message Authentication Code}
	\acro{ML}[ML]{Machine Learning}
	\acro{MTU}[MTU]{Maximum Transmission Unit}
	\acro{MOST}[MOST]{Media Oriented System Transport}
	\acro{NAD}[NAD]{Network Anomaly Detection}
	\acro{NADS}[NADS]{Network Anomaly Detection System}
	\acroplural{NADS}[NADSs]{Network Anomaly Detection Systems}
	\acro{NC}[NC]{Network Calculus}
	\acro{OEM}[OEM]{Original Equipment Manufacturer}
	\acro{OMG}[OMG]{Object Management Group}
	\acro{OTA}[OTA]{Over-the-Air}
	\acro{P4}[P4]{Programming Protocol-independent Packet Processors}
	\acro{PCAP}[PCAP]{Packet Capture}
	\acro{PCAPNG}[PCAPNG]{PCAP Next Generation Dump File Format}
	\acro{PCP}[PCP]{Priority Code Point}
	\acro{PKI}[PKI]{Public Key Infrastructure}
	\acro{PSFP}[PSFP]{Per-Stream Filtering and Policing}
	\acro{RAP}[RAP]{Resource Allocation Protocol}
	\acro{RC}[RC]{Rate-Constrained}
	\acro{REST}[ReST]{Representational State Transfer}
	\acro{RPC}[RPC]{Remote Procedure Call}
	\acro{RTPS}[RTPS]{Real-time Publish-Subscribe}
	\acro{SaaS}[SaaS]{Software-as-a-Service}
	\acro{SD}[SD]{Service Discovery}
	\acro{SDN}[SDN]{Software-Defined Networking}
	\acro{SDN4CoRE}[SDN4CoRE]{Software-Defined Networking for Communication over Real-Time Ethernet}
	\acro{SecVI}[SecVI]{\textit{Security for Vehicular Information}}
	\acro{SOA}[SOA]{Service-Oriented Architecture}
	\acroplural{SOA}[SOAs]{Service-Oriented Architectures}
	\acro{SOA4CoRE}[SOA4CoRE]{Service-Oriented Architecture for Communication over Real-Time Ethernet}
	\acro{SOME/IP}[SOME/IP]{\textit{Scalable service-Oriented MiddlewarE over IP}}
	\acro{SPOF}[SPOF]{Single Point of Failure}
	\acro{SR}[SR]{Stream Reservation}
	\acro{SRV}[SRV]{Service}
	\acro{SRP}[SRP]{Stream Reservation Protocol}
	\acro{SVM}[SVM]{Support Vector Machine}
	\acro{SVCB}[SVCB]{Service Binding}
	\acro{SW}[SW]{Switch}
	\acroplural{SW}[SWs]{Switches}
	\acro{TAS}[TAS]{Time-Aware Shaper}
	\acro{TCP}[TCP]{Transmission Control Protocol}
	\acro{TDMA}[TDMA]{Time Division Multiple Access}
	\acro{TN}[TN]{True Negative}
	\acro{TP}[TP]{True Positive}
	\acro{TLS}[TLS]{Transport Layer Security}
	\acro{TSN}[TSN]{Time-Sensitive Networking}
	\acroplural{TSN}[TSN]{Time-Sensitive Networks}
	\acro{TSSDN}[TSSDN]{Time-Sensitive Software-Defined Networking}
	\acro{TT}[TT]{Time-Triggered}
	\acro{TTE}[TTE]{Time-Triggered Ethernet}
	\acro{TTL}[TTL]{Time to Live}
	\acro{UDP}[UDP]{User Datagram Protocol}
	\acro{UN}[UN]{United Nations}
	\acro{QoS}[QoS]{Quality-of-Service}
	\acro{V2X}[V2X]{Vehicle-to-X}
	\acro{WS}[WS]{Web Services}
	\acro{ZC}[ZC]{Zone-Controller}
	\acroplural{ZC}[ZCs]{Zone-Controllers}
	\acro{ZSK}[ZSK]{Zone Signing Key}
\end{acronym}
\section{Introduction}
\label{sec:intro}

Vehicles comprise a distributed system of software-defined and hardware-controlled functions that enhance vehicle performance, safety, and comfort. 
Software innovations contribute substantially to the car value~\cite{c-hsijr-21,bmle-cssjr-23} with vehicular services that can be dynamically orchestrated based on customer-selected configurations and available resources~\cite{lbdll-eaujr-24,hmks-stsnv-23}.
Online connected vehicles open innovation options further and enable \acp{ADAS}, which interact with infrastructure services and environmental data via \ac{V2X} communication.
However, this desired flexibility combined with global Internet exposure extends the attack surface.  
Remote manipulation attacks by third parties~\cite{cmkas-ceaas-11,zprsk-tmajr-22,u-2gajr-24} were already observed in the wild~\cite{mv-reupv-15,bbcv-yctjr-25}.

Vehicles are long-lasting products that require equally durable security solutions, which must be maintained throughout their lifetime---as reflected in automotive standards, \eg ISO/SAE 21434~\cite{iso-sae-21434}.
Nevertheless, many automotive protocols, originally designed for closed systems, completely lack security features~\cite{yl-sacjr-20,zprsk-tmajr-22,ii-fadjr-24}.
This includes the \ac{SOME/IP}~\cite{a-spsjr-24,a-ssdjr-24}, the current choice of the \ac{AUTOSAR} for publish-subscribe communication, which has been shown vulnerable to various attacks~\cite{ii-fadjr-24,zlkk-assjr-21}.
Previous attempts to secure the protocol~\cite{irrsv-ssvjr-20,zlkk-assjr-21,myzzc-ascjr-22,lcl-pscjr-23} have focused on tailored, tightly integrated solutions for message encryption, authentication, or authorization. 
These solutions reinvent crypto protocols 
and build on certificates pre-shared across all communication partners.

\begin{figure}
    \centering
    \includegraphics[width=1\linewidth, trim=24pt 22pt 24pt 22pt, clip=true]{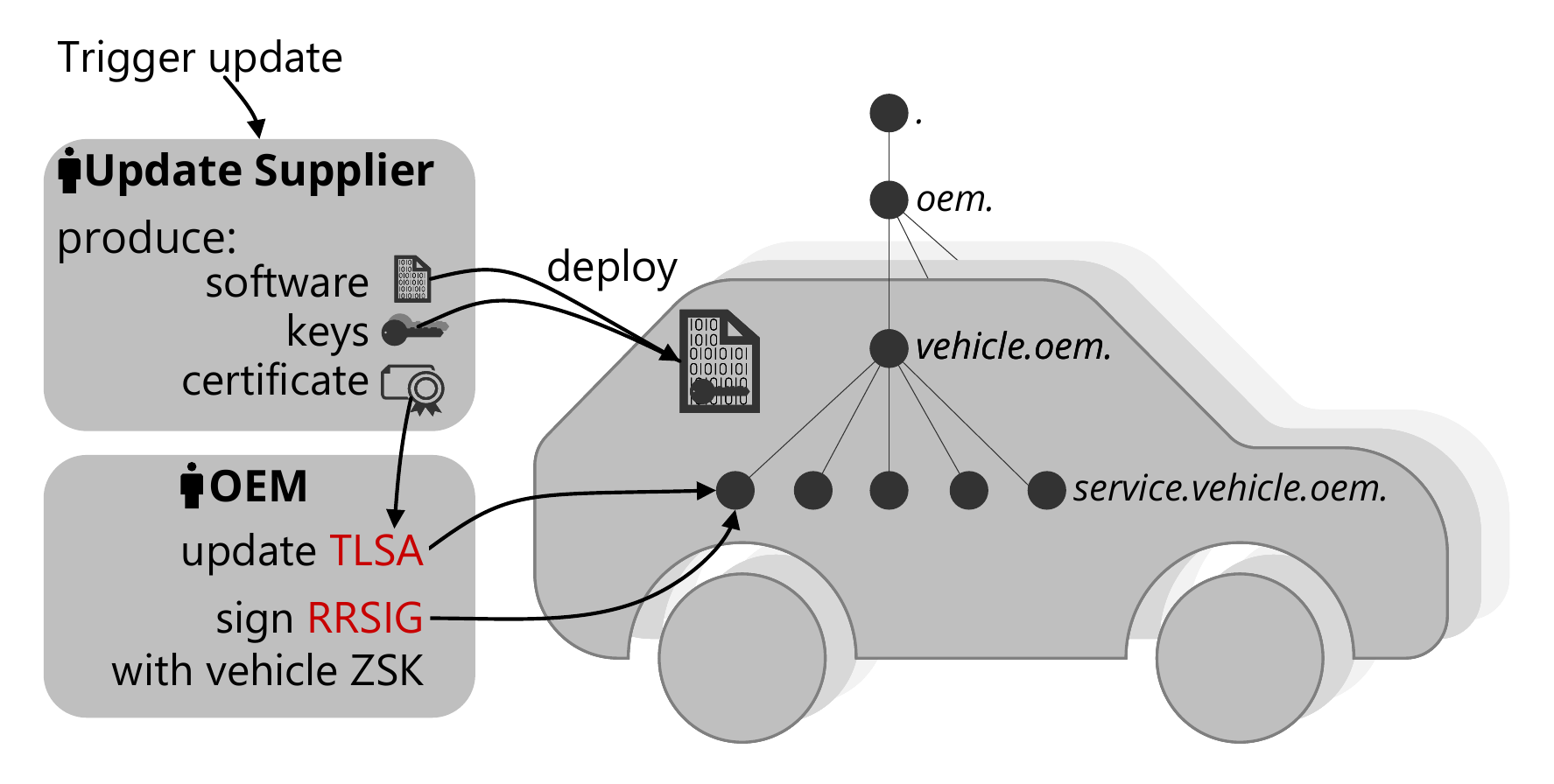}
    \caption{
    DNSSEC-based credential management for automotive services that decouples software signing by the supplier from record signing by the OEM.
    }
    \label{fig:cert_management}
\end{figure}

The complexity of in-vehicle services demand for \one a scalable and robust key management to accommodate \two secure authentication and authorization of services and clients. 
In the automotive ecosystem, the \ac{OEM} is responsible for the overall security lifecycle of a car. 
While suppliers retain accountability for the security of their individual components, the \ac{OEM} is ultimately held liable for the system integrity, in particular in public perception.
This demands coordination between various stakeholders, including tier-1 suppliers for critical components, specialized software vendors, and aftermarket add-on contributors. 
The heterogeneity of this ecosystem poses severe challenges on key provisioning and robust credential management.

In this paper, we address these challenges by leveraging the established \ac{DNSSEC}~\cite{RFC-2535-custom} ecosystem.
\cref{fig:cert_management} illustrates our approach that decouples the software signing (supplier) from credential deployment (\ac{OEM}).
The supplier signs its product and a corresponding certificate, while the \ac{OEM} signs the deployment of the certificate in a DNSSEC TLSA record using independent credentials. 
A delegated \ac{DNSSEC} zone per vehicle cached within the car enables offline credential validation and supports mid-term scenarios without Internet access.

Based on this DNSSEC credential management, we derive a security layer for automotive service architectures. 
The DNS offers a scalable, distributed database that stores relevant data for service endpoint security and supports remote record updates and key rollover within its resolver infrastructure.
\ac{DNSSEC} provides verification of data integrity and authenticity for all of its records.  
\ac{DNSSEC} deployment and maintenance on the Internet are supported by robust tools and guidance, adaptable to private DNS infrastructure where the \ac{OEM} manages the entire \ac{DNSSEC} environment---including the root/trust anchor---within an independent namespace.

The vehicular use-case for DNSSEC introduces novel challenges over typical Internet DNSSEC, especially with regards to long-lived in-car credentials, supply-chain aspects, intermittent connectivity, and safety-critical operation which we address in this work.
In detail, our main contributions read:\vspace{-2pt}
\begin{itemize}
	\item We derive automotive security requirements~(\S \ref{sub:requirements}) and show how the DNSSEC ecosystem can comply with them~(\S \ref{sec:DNSSEC-Case}).
	\item We introduce DNSSEC-based authentication and authorization for publishers \textit{and} subscribers in the automotive system (\S \ref{sec:concept}).
    \item We propose secure key lifecycle management and deployment that comply to heterogeneous trust and operative requirements of the automotive domain, such as offline validation (\S \ref{sec:application}). 
    \item We integrate our approach with SOME/IP (\cref{sec:someip}) and evaluate its security (\S \ref{sec:security}) and performance (\S \ref{sec:evaluation}) to demonstrate its effectiveness in practice. 
\end{itemize}

\section{The Problem of Securing Automotive Services and Related Work}
\label{sec:background}

Modern vehicular environments follow the \ac{SaaS} paradigm for flexibility, service availability, customization, and upgradeability~\cite{c-hsijr-21,bmle-cssjr-23,a-naejr-24,b-fsnjr-25}.
Optional features, agile deployments, and frequent updates benefit from a loosely coupled \ac{SOA}~\cite{lbdll-eaujr-24}. 

\begin{figure}
    \centering
    \includegraphics[width=\linewidth,trim=46pt 50pt 26pt 50pt, clip=true]{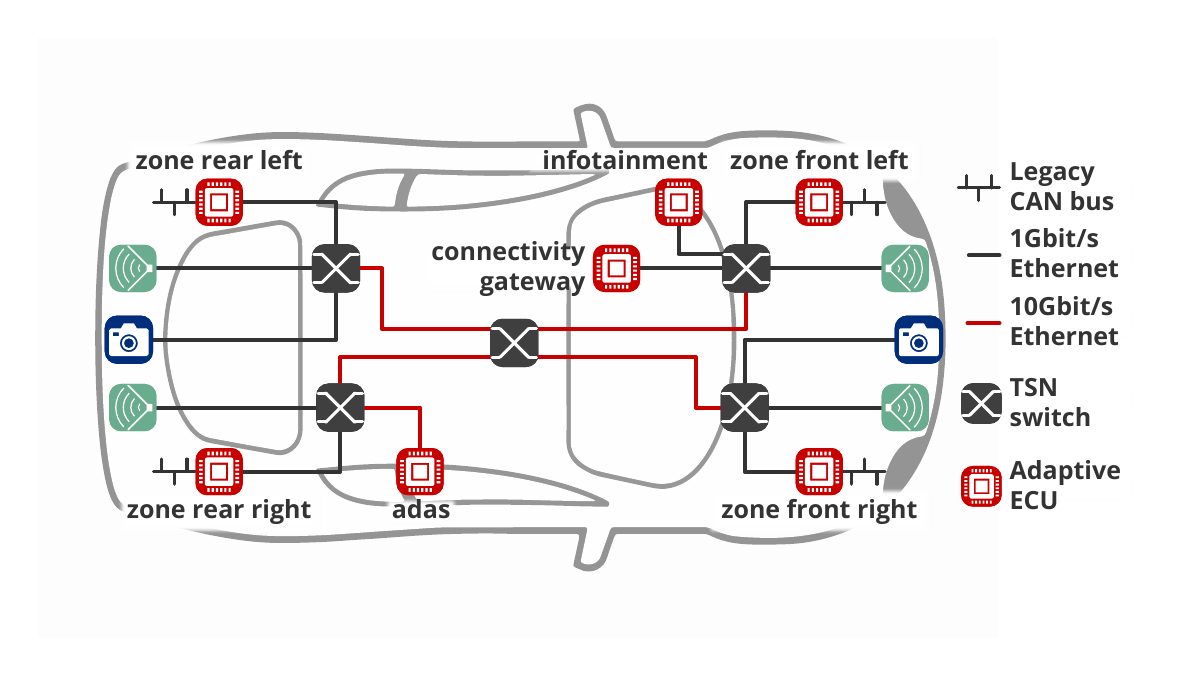}
    \caption{In-Vehicle Network (IVN) in a zone topology with adaptive multi-service compute units and a unified Ethernet backbone.}
    \label{fig:ivn_topology}
\end{figure}

This transformation is reflected on the infrastructure level by a shift to multi-service \acp{HPC}~\cite{bmle-cssjr-23}.
A high-speed Ethernet backbone~\cite{mk-ae-21}---real-time enhanced by \ac{TSN}~\cite{ieee8021q-22}---replaces the traditional \ac{CAN} domain topology with dedicated \acp{ECU}. 
\cref{fig:ivn_topology} illustrates such a future \ac{IVN} interconnecting \acp{HPC} in physical zones~\cite{wtm-avnjr-21,hmks-stsnv-23}, as built into new car models, \eg by Audi~\cite{a-naejr-24} and BMW~\cite{b-fsnjr-25}. 

Middleware layers integrate the heterogeneous multi-supplier \ac{SOA} via dynamic service discovery, instantiation, and migration using a publish-subscribe model over UDP/IP multicast~\cite{hmks-stsnv-23}.
Prominent candidates include \ac{DDS}\cite{omg-dds-15} and \ac{SOME/IP}\cite{a-spsjr-24,a-ssdjr-24}, the latter specifically tailored for automotive needs.
We prototype our approach using \ac{SOME/IP} due to widespread automotive adoption~\cite{a-spsjr-24, iks-aiejr-22} but remain protocol-agnostic and adaptable to \ac{DDS}. 

\subsection{Security Objectives, Threats, and Protection}
The safety of modern cars is tightly bound to the security of the various hardware and software components. 
Cyber attacks~\cite{mv-reupv-15,u-2gajr-24} endanger passenger safety, compromise user privacy, and pose financial and reputational risks to \acp{OEM}~\cite{zprsk-tmajr-22}.
The main security objectives for the \ac{IVN} are service \textit{availability}, data \textit{integrity}, and endpoint \textit{authenticity}~\cite{mbzls-ssajr-18,dzjal-adijr-20,hmks-stsnv-23}. 

The vehicular attack surface spans different local and remote interfaces~\cite{cmkas-ceaas-11,dzjal-adijr-20}, 
the latter extend the surface to global access and remote attacks. 
Threats may propagate across the \ac{IVN}~\cite{dzjal-adijr-20,zprsk-tmajr-22} and the centralized, shared architecture may amplify the vulnerability of safety-critical functions. 
Potential attackers range from organized crime over hacktivists and nation-state actors to the vehicle owners~\cite{mbzls-ssajr-18,zprsk-tmajr-22}, each of which can act alone or in groups at different professional levels~\cite{iso-sae-21434}. 
Related work has exposed severe vulnerabilities in automotive middleware including timing-based \ac{DoS} attacks, spoofing of discovery messages, and unauthorized access~\cite{ii-fadjr-24,zlkk-assjr-21}.

Existing automotive threat collection, such as the automotive threat matrix (ATM)~\cite{atm}, cover only a limited subset of in-car communication risks and lack the structured rigor of established threat models. 
We aim at comprehensively addressing network-related threats, which are prevalent in Ethernet-based \acp{IVN}, and hence opt for a standard threat analysis framework for communication systems to evaluate our proposed authentication and authorization scheme.
The STRIDE~\cite{hl-sdljr-06} threat model covers \textbf{S}poofing, \textbf{T}ampering, \textbf{R}epudiation, \textbf{I}nformation disclosure, \textbf{D}enial of service, and \textbf{E}levation of privilege, which we complement with automotive-specific attack scenarios~\cite{ii-fadjr-24,zlkk-assjr-21,atm}. 

Security measures span different layers, and include attack prevention, detection, and mitigation~\cite{tw-avsjr-16}.
Prevention uses firewalls~\cite{psz-hcdjr-17}, network flow separation~\cite{hmks-stsnv-23}, and state-aware defense~\cite{xllzl-ssajr-22a}. 
Intrusion detection systems identify ongoing attacks~\cite{rmwh-sadjr-18,mhlks-fsaad-24}.
Message integrity can be ensured with authentication codes and encryption for \acp{IVN}~\cite{mpmsl-sanjr-17,hl-rscan-18,irrsv-ssvjr-20}. 
Secure boot~\cite{sksb-sbrjr-20} and \ac{OTA} updates~\cite{mns-stajr-22} protect against unauthorized software modifications and patch vulnerabilities.
The automotive \ac{SOA} requires robust measures that prevent manipulation of service announcements and \mbox{(un-)}subscriptions~\cite{zlkk-assjr-21,ii-fadjr-24}.
In this work, we focus on authentication and authorization of service discovery and invocation -- essential to protect the \ac{SOA} and orthogonal to existing solutions. 
We summarize the \textbf{requirements for authentication and authorization of vehicular services:}

\label{sub:requirements}
\begin{description}[style=unboxed,leftmargin=0.33cm]
    \item[Service continuity:] The critical driving functions of a vehicle must remain operational and safe under all conditions, independent of remote connectivity. 
    This resilience must be sustained under continuous software updates and security lifecycle management.
    \item[Long-term, agile credential management:] Procedures for robust signing, certificate renewal, and key rollover must manage heterogeneous vehicle fleets across their lifecycle.
    \item[Replicable, cacheable service credentials:] Service credentials must be self-consistent and locally verifiable---independent of any (remote) attestation service. Content should be secured on the object level, which must not depend on content location or suppliers. 
    \item[Manageability:] Key management and distribution should be efficient and confined to trustees.
    \item[Scalability:] Solutions must scale with a growing number of cars, services, and backend systems.
    \item[Interoperability and distribution:] Security solutions must cater for distributed suppliers. Well established standards should ensure protocol interoperability across stakeholders. 
    \item[Performance:] While the service discovery itself is neither safety- nor time-critical, the authentication and authorization process should meet latency limits ($<\SI{200}{\milli\second}$~\cite{hmks-stsnv-23}).
\end{description}

\subsection{Service Authentication and Authorization}
In a car, the trust relations between the manufacturer and the service suppliers must map to cryptographic trust relations to secure service interoperability. 
Service authentication can establish this trust by verifying endpoints via certificates derived from a common trust anchor. 
Endpoints can confirm authenticity via a challenge-response scheme to prove ownership of the corresponding private key.
Legacy \acp{IVN} rely on protocol-specific authentication and authorization with limited interoperability. 
\ac{CAN} bus security approaches, for instance, are tailored to shared media and face bandwidth and frame size constraints~\cite{mpmsl-sanjr-17,wbbzw-cmsjr-25}. 
Constrained \acp{ECU} hinder robust security solutions~\cite{zkss-utsjr-17}. 
Modern \ac{HPC} platforms are more powerful and allow for advanced security mechanisms~\cite{zkss-utsjr-17,hmks-stsnv-23}.

Established security protocols, such as MACsec, IPSec, and (D)TLS, have been proposed to protect service middlewares that lack integrated security features~\cite{lzkkv-spejr-24,zkss-utsjr-17}.
However, these protocols face limitations in the automotive \ac{SOA}~\cite{yl-sacjr-20,irrsv-ssvjr-20}. 
MACsec (layer 2) authenticates physical devices, while IPSec (layer 3) operates between IP endpoints. 
Neither authenticates services in the presence of multi-service controllers. 
Even though (D)TLS operate on the service level, they cannot serve multicast channels. 
Multicast accounts for 75\% of in-vehicle control flows~\cite{hmks-stsnv-23} and carries the service discovery~\cite{irrsv-ssvjr-20}. 
This leaves the \ac{SOA} vulnerable to timing attacks with false service announcements, which can disrupt legitimate instances~\cite{ii-fadjr-24}.

Previous work has addressed \ac{SOME/IP} vulnerabilities~\cite{ii-fadjr-24,zlkk-assjr-21} by message encryption, authentication, and access control.
These approaches depend on custom handshakes and external authentication servers. 
Iorio~\etal\cite{irrsv-ssvjr-20} add a challenge-response handshake for authenticating, authorizing, and encrypting \ac{SOME/IP} payloads.
Zelle~\etal\cite{zlkk-assjr-21} secure service discovery multicasts using \ac{ECU} authentication and message validation via an authorization server.
Ma~\etal\cite{myzzc-ascjr-22} employ \acp{MAC} to prevent tampering and use a domain controller for mutual \ac{ECU} authentication.
Lee~\etal\cite{lcl-pscjr-23} introduce a ticket system, for which an authentication server grants subscription permissions.
Neither solution explicitly authenticates nor authorizes service discovery or subscription, which is essential for service integrity in cars.

In the X.509 \ac{PKI}~\cite{RFC-5280}, trusted \acp{CA} certify authenticity of Internet applications but 
 the public \ac{CA} model allows any trusted \ac{CA} to issue certificates for any resource, potentially leading to attestation collisions~\cite{RFC-6698}.
The DDS standards define a tailored authentication extension based on a \ac{PKI}~\cite{omg-dds-security-18} and thus faces the same issue.
In the automotive sector, a trusted supplier could issue certificates for any service in any car and thus enable cross-vehicle escalation of privileges. 

Previously, we proposed~\etal\cite{mhmks-asasd-23} DNSSEC-based authentication for publisher validation, replacing the distributed discovery mechanism of SOME/IP with DNSSEC lookups. 
The discovery, however, serves broader purposes that include signaling service availability, mobility, and service instance selection.
In this work, we do not replace in car service discovery by DNS lookups but rather use the DNSSEC for secure certificate provisioning during subscription establishment.
Critical gaps remain in client authentication, access control, and certificate management---issues that this work addresses to provide a comprehensive authentication and authorization framework for automotive services.

\subsection{Security Credential Management}
\ac{OTA} software updates are increasingly adopted in the automotive domain, yet security credential updates received little attention.
Automotive security extensions~\cite{irrsv-ssvjr-20,zlkk-assjr-21,myzzc-ascjr-22,lcl-pscjr-23} rely on pre-deployed certificates and static access control lists that cause two critical conflicts in the supply chain: 
\one \acp{OEM} must pre-configure all credentials of potential peers during manufacturing, and face cascading update demands for related services when adding or updating a service.
\two Suppliers cannot independently update credentials of their components, even if expired or compromised. 

Static or pre-deployed certificates pose significant risks over the long lifetime of a car, contradicting the NIST recommendation to limit authentication key lifetimes to one or two years~\cite{n-rkmjr-20}.
Yet efficient, secure key rotation and distribution processes are missing to protect the vehicle over its lifetime.
Established certificate management, such as the Web\ac{PKI}, do not cater for the supplier-\ac{OEM} relationships, but require delegation of trust to either the suppliers or the \ac{OEM}. 

Credential bootstrapping for device onboarding is addressed by several standards, \eg
FDO~\cite{fido-22}, 
SZTP~\cite{RFC-8572}, or 
BRSKI~\cite{RFC-8995}. 
These provision initial security credentials during deployment that enable device authentication, but do not cover application authentication or certificate lifecycle management. 
We consider device onboarding orthogonal to our focus on service authentication and certificate management and aim to provide a framework for mutual authentication that eliminates cascading updates when client or server credentials change.

Recently, P\"ullen~\etal\cite{pfclk-spajr-24} proposed credential management via a central authentication server that distributes access tokens across \acp{ECU}. 
The solution follows a customized top-down procedure for deploying credentials to this central server, which involves application designers in credential management, which may impact robustness, interoperability, and scalability.
Such approaches couple the in-car trust chain with the software supply chain, which weakens the security model.

\customparagraph{The DNSSEC ecosystem.}
The DNS 
has evolved into a rich ecosystem of information, application, and tooling. 
Various DNS records can be transparently accessed, cached, and delegated, while \ac{DNSSEC}~\cite{RFC-2535-custom} protects their integrity and authenticity via signature records (RRSig). 
As such, DNSSEC is interoperable and performant on Internet scale and caters for state-of-the-art cryptography.
20 years of operational experience have proven robust key management~\cite{otsw-bktf1-22}. 

\ac{DANE}~\cite{RFC-6698} and \ac{DANCE}~\cite{draft-ietf-dance-architecture} build on \ac{DNSSEC} to store and sign keys and certificates in TLSA records.
They bind a certificate to a named service endpoint within the DNS hierarchy, establishing a chain of trust from the root zone to a service certificate (see \cref{fig:dnssecsigning}). 
This avoids the well-known attestation collisions problem present in the WebPKI~\cite{RFC-5280}, and, in particular, allows for independent signing of the service resource by its certificate owner and the service deployment by the owner of the DNS TLSA record. 
We will use this double signing to decouple third-party service provider credentials from those of the \ac{OEM} for secure in-car service deployment.

The security of \ac{DNSSEC}, \ac{DANE}, and \ac{DANCE} has been thoroughly discussed in the IETF standardization~\cite{RFC-2535-custom,RFC-6698,RFC-7671,RFC-7673,draft-ietf-dance-architecture} and analyzed extensively in research~\cite{otsw-bktf1-22,lamrk-uhdms-22,mtwhc-rrryr-19,lgrkc-lcsde-20,thlsw-dccdi-24,k-vddjr-14,ldhbv-ffejr-23,zskbl-pdcjr-25,mjhor-rpqjr-20}. 
The DNSSEC ecosystem has proven robust and resilient over many years on the Internet, and the same established methods and tools can be used for private namespaces of OEMs or cars.

\section{The Case for DNSSEC, DANE, and DANCE}
\label{sec:DNSSEC-Case}
In this work, we address service authentication and authorization with integrated certificate management in cars. 
Our approach unifies credential management across automotive service middlewares using established \ac{DNSSEC} mechanisms that provide a scalable, secure, and standardized solution.

\begin{figure}
    \centering
    \includegraphics[width=0.75\linewidth, trim=28pt 14pt 18pt 18pt, clip=true]{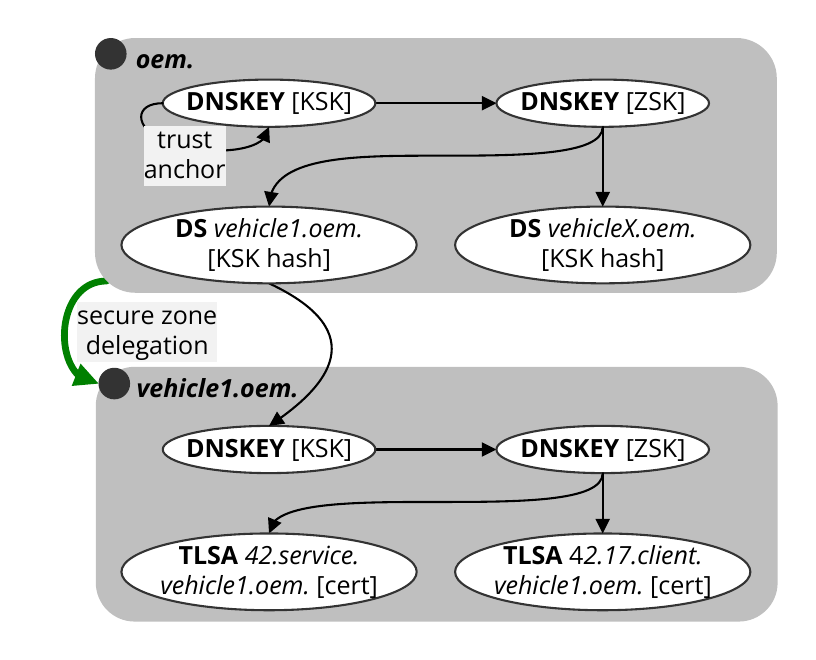}  
    \caption{
        DNSSEC chain of trust showing the OEM-controlled zone as a trust anchor with secure per car zone delegation, which signs service and client certificates in TLSA records via RRSIG records.
    }
    \label{fig:dnssecsigning}
\end{figure}

\customparagraph{Leveraging (private) DNSSEC infrastructure.}
The distributed nature of \ac{DNSSEC} supports zone delegation and regional replication.
An OEM can manage independent namespaces within private infrastructure, including the DNSSEC root (trust anchor), which eliminates dependency on public DNS infrastructure. 
RFC 6394~\cite{RFC-6394} covers DANE use cases that match our requirements for IVN service authentication.
\cref{fig:dnssecsigning} depicts a \ac{DNSSEC} deployment for an \ac{OEM} and a car, featuring TLSA records for services and clients. 
The \ac{KSK} at the root serves as a trust anchor for all resolvers and signs the \acp{ZSK} that in turn sign individual record sets. 
Delegation Signer (DS) records cryptographically link parent and child zones.
This ecosystem facilitates secure key management and mutual authentication along the name hierarchy, with the \ac{OEM} maintaining control over the backend and vehicle service records.

\customparagraph{Isolating vehicle credentials.}
The \ac{OEM} can delegate a \ac{DNSSEC} zone for each vehicle to ensure isolated credentials for each car. 
Within this zone, dedicated \acp{ZSK} sign the \ac{DANE} records for all service and client credentials of a car (see \cref{fig:dnssecsigning}).
On-board recursive resolvers cache these records (with long TTLs) for secure offline operation, which is critical in remote or disconnected environments. 
DNSSEC record updates and key renewals are only performed when the car is at rest.
This framework delivers a robust and scalable management for automotive service credentials.

\customparagraph{Decoupling supplier security and in-vehicle trust.}
In the automotive domain, \acp{OEM} can control \ac{DNSSEC} deployment across an entire fleet and all backend systems, which ensures compatibility across clients and servers.
Our design decouples the supply of service credentials from the in-vehicle chain of trust provided by the \ac{OEM} (see \cref{fig:cert_management}), offering a solution for the multi-supplier automotive ecosystem without key escrow.
Suppliers can independently generate software updates with their security credentials, while the \ac{OEM} maintains the DNS hierarchy and guarantees namespace integrity.

\customparagraph{Middleware-independent solution.}
We design protocol-independent authentication and authorization for automotive services, enabling interoperability, longevity, and scalability based on \ac{DNSSEC} standards.
To demonstrate effectiveness, we integrate our solution with \ac{SOME/IP}, which is widely adopted in the automotive domain as the current choice of the AUTOSAR consortium~\cite{a-spsjr-24, iks-aiejr-22}. 
\ac{SOME/IP} lacks robust security mechanisms. 
Our design remains protocol-agnostic and adaptable to alternative middlewares such as \ac{DDS}. 
We incorporate a challenge-response scheme into \ac{SOME/IP} service discovery, securing service announcements and subscriptions.

In the following sections, we present this integration in detail and evaluate how the DNSSEC ecosystem can fulfill the requirements for comprehensive in-car service authentication and authorization. 
We demonstrate the feasibility of our proposal in a realistic automotive scenario.

\section{DNSSEC-based Authentication and Authorization for In-Car Services} 
\label{sec:concept}

\begin{figure}
    \centering
    \includegraphics[width=0.85\linewidth, trim=44pt 20pt 20pt 32pt, clip=true]{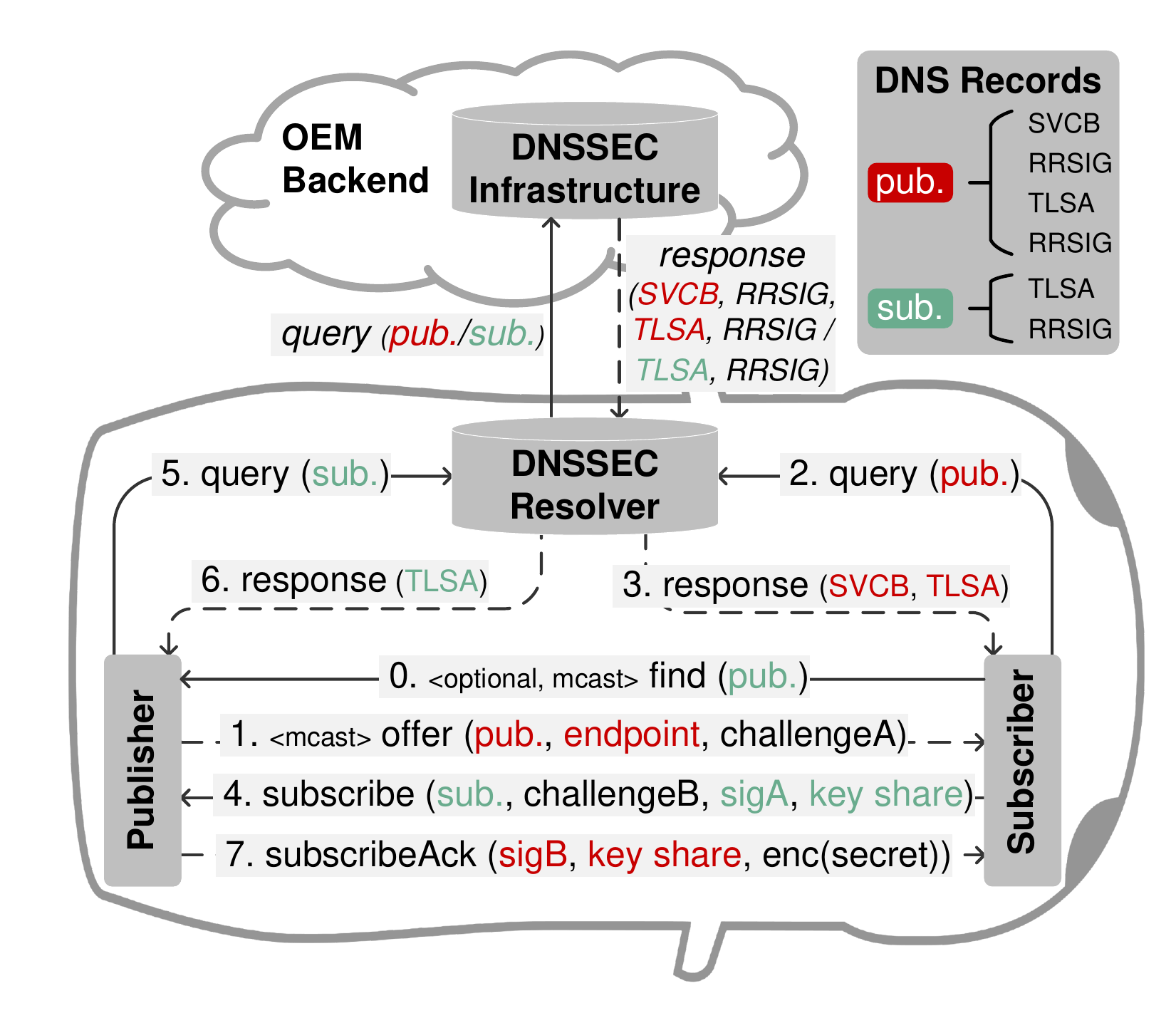}  
    \caption{
        DNSSEC-based authentication and authorization for automotive publish-subscribe systems. 
        \textcolor{CoreRed}{Publisher} and \textcolor{CoreGreen2}{subscriber} information are color coded. 
        The remote DNSSEC infrastructure may be disconnected while driving. 
    }  
    \label{fig:concept}
\end{figure}

\cref{fig:concept} displays the key components of the automotive publish-subscribe system running \ac{SOME/IP}.
A local \ac{DNSSEC} resolver in the car caches all records of its zone, which it pre-loads from the backend and validates. 
This design ensures offline functionality during normal vehicle operations.
 
During the service discovery at startup, publishers and subscribers query the local resolver and authenticate each other using a challenge-response handshake.
Thereafter all sessions are established and the vehicle can drive safely without \mbox{(re-)}configurations.

\customparagraph{Acquiring valid endpoint information.}
\label{sec:endpoint}
In \ac{SOME/IP}, subscribers can trigger service discovery via optional find messages.
Publishers respond to these requests or periodically offer their services to signal availability. 
After the discovery, subscribers query the local \ac{DNSSEC} resolver, which provides the service data encoded in \ac{SVCB} records~\cite{RFC-9460}. 
We selected \ac{SVCB} for its flexibility and extensibility, as it supports various \acp{ALPN} (\eg HTTPS over TCP) and includes the necessary address, port, and protocol-specific parameters. 
A single query may return multiple \ac{SVCB} records, for instance, when a service supports multiple protocols or operates redundant instances. 
Subscribers cross-references the \ac{DNSSEC}-protected \ac{SVCB} data with the published offer to validate the endpoint information before issuing a subscription.

\customparagraph{Authenticating publishers and subscribers.}
\label{sec:authentication}
\ac{DANE}~\cite{RFC-6698} and \ac{DANCE}~\cite{draft-ietf-dance-architecture} bind certificates to endpoints and clients via TLSA records.
We integrate a challenge-response mechanism into the existing publish-subscribe handshake. 
This design authenticates service discovery, subscription, and data exchange in a single step, eliminating the need for a separate post-subscription handshake (as required by traditional (D)TLS-based solutions, \eg~\cite{irrsv-ssvjr-20}).
We describe our \ac{SOME/IP} implementation in the \cref{sec:someip}.

The authentication workflow begins with publishers embedding an authentication challenge-nonce in their offer.
Subscribers sign the challenge with their private key and include it in the subscribe message. 
In turn, subscribers can send another challenge-nonce in subscribe messages to request authentication from publishers, that return a signed response in the subscription acknowledgment.
Both parties query the local resolver for TLSA records of their peers based on the name already learned during discovery.
\codeword{DANE-EE} (End Entities) certificate usage eliminates dependence on an external \ac{CA} by anchoring trust exclusively in the \ac{DNSSEC} chain.
TLSA records contain a full certificate or a hash against which the presented signatures are verified.
Again, a query may return multiple records, which allows for certificate migration.

For performance, subscribers may fetch \ac{SVCB} and TLSA records in parallel, ignoring the latter if the SVCB record is insecure~\cite{RFC-7673}.
If TLSA records are missing or insecure, the connection may proceed without cryptographic assurance---subject to implementation-specific policies~\cite{RFC-6698}.

\customparagraph{Name-based authorization for services.}
\label{sec:authorization}
Traditional authorization mechanisms often rely on access control lists or complex infrastructure, which provoke cascading updates and error-prone maintenance. 
Instead, we reuse existing, protected TLSA record names for service authorization, leveraging the same \ac{DNSSEC} trust chain without additional infrastructure or cryptographic overhead.
A valid \ac{DNSSEC}-secured TLSA record confirms the legitimacy of an offer or subscription.

Authorization by name supports fine-grained policies but remains simple to manage and deploy.
Publisher record names follow the format \textit{service.domain.vehicle.oem.}, with prepended protocol information. 
Subscribers authorize publishers through exact name matching of \ac{DNSSEC}-secured TLSA records.
To support scoped access policies (\eg per domain or vehicle-wide), we introduce a hierarchical naming scheme for subscriber certificates with optional fields: \textit{\_proto-client.sub[.pub][.domain][.vehicle].oem.}
Publishers define accepted scopes based on application needs, 
\eg requiring service-specific names for tighter authorization.

\customparagraph{Session establishment and key exchange.}
\label{sec:keyexchange}
We establish session keys using \ac{DH} key exchange~\cite{dh-ndcjr-76}. 
The subscriber includes its public share in the subscription, and the publisher in the subscription acknowledgement. 
Both parties can then compute the symmetric session key. 

For multicast endpoints, the session key must be agreed upon between multiple parties, for which several schemes (including DH) exist.  
In our prototype, the publisher acts as key sponsor that individually supplies a symmetric group key to its subscribers via the established secure channels. 

The symmetric key can be used to encrypt and decrypt unicast and multicast messages, which ensures confidentiality and integrity.
The publisher updates the session key when required using its secure channels.
Furthermore, secure and fast session resumption mechanisms, \eg following DTLS, may reduce overhead and improve performance, but are beyond the scope of this work.

\section{In-Vehicle Deployment and Operation}
\label{sec:application}
We now discuss how this scheme addresses the specific demands of in-vehicle operations and the automotive constraints. 
Safety and security are critical for cars, with \acp{OEM} bearing responsibility for the entire vehicle lifecycle, which subsumes third-party components. 
Regulators mandate repair and maintenance in independent workshops, which requires simple, standard interfaces for securing component updates.

\subsection{Certificate Management}
\label{sec:cert_management}

Effective certificate management is a cornerstone of the proposed solution.
The \ac{OEM} oversees the entire fleet while many suppliers provide updates of hardware and software.
Our approach (see \cref{fig:cert_management}) separates the software signing (by suppliers) from the record signing (by the \ac{OEM}), which ensures the \ac{OEM} retains full control over the in-car deployment via the DNS.
All credentials remain with their owners without involving a third-party (\eg a \ac{CA}) or key escrow.
Leveraging \ac{DNSSEC} for credential storage and distribution streamlines processes and enhances security with robust tools, best practices, and in-depth security analysis~\cite{otsw-bktf1-22,mtwhc-rrryr-19}.

\begin{figure}
    \centering
    \includegraphics[width=\linewidth, trim=18pt 20pt 18pt 18pt, clip=true]{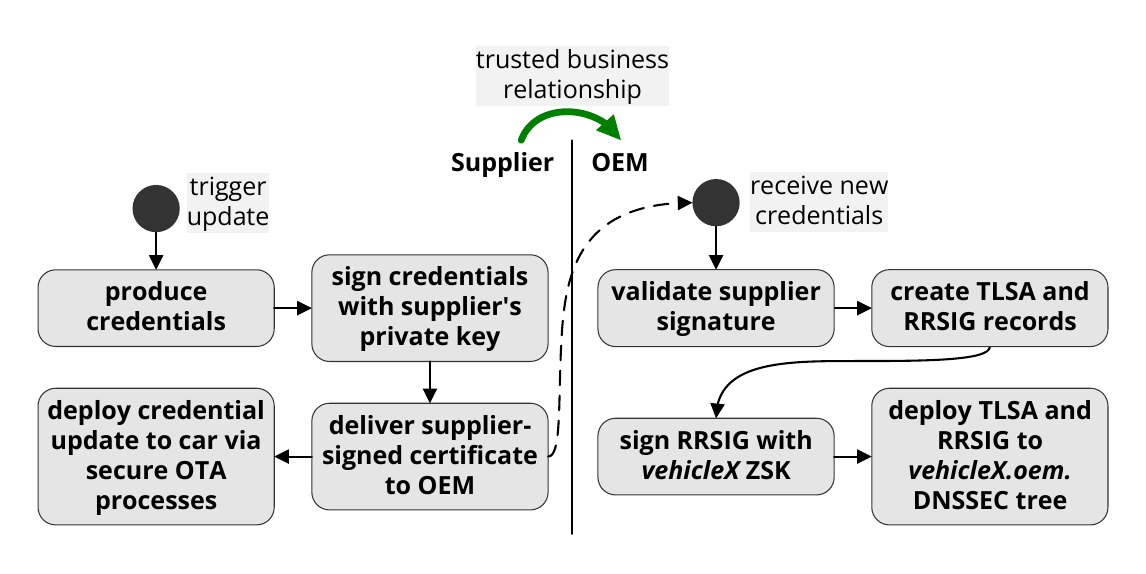}  
    \caption{
        Decoupled credential signing: supplier-signed credentials and \ac{OEM}-signed \ac{DNSSEC} deployment per vehicle \ac{ZSK}.
    }
    \label{fig:credentialsigning}
\end{figure}

\customparagraph{Service credential rotation} is essential. NIST recommends updating authentication keys every one or two years~\cite{n-rkmjr-20}. 
\cref{fig:credentialsigning} shows the credential update mechanism with decoupled supplier and OEM signing roles. 
The supplier produces software binaries, key pairs, and corresponding certificates---independent updates of credentials and software are possible.
The supplier then signs all produced artifacts with its private key, delivers the public certificate to the OEM, and provides the update image for the respective car. 
Secure \ac{OTA} update procedures are readily used to deploy software updates to modern cars~\cite{mns-stajr-22}. 
In this context, we assume that the supplier and the \ac{OEM} have means to establish a secure, trusted channel following contracts and business relationships.
 
The \ac{OEM} validates attached signatures against known supplier credentials, creates a TLSA record from the service certificate, and signs it in an RRSIG record using the vehicle \ac{ZSK} (see \cref{fig:dnssecsigning}).
TLSA and RRSIG records are deployed to the DNS tree, from which they can be retrieved and validated by the local in-car resolver.
Multiple TLSA records for the same service can coexist and allow for overlapping credentials during a rotation.
A secure \ac{OTA} process updates the car software and stores its private key locally, \eg in a \ac{HSM}.

The inherent flexibility of \ac{DNSSEC} permits additional zone delegations, facilitating the integration of third-party components, \eg by car rental companies.
To support the right to repair, \acp{OEM} should provide tooling that enables suppliers and independent garages to securely update software, credentials, and trigger renewal of a \ac{DNSSEC} record.

\customparagraph{Certificate uniqueness.}
In contrast to related work, each car in its delegate zone (\eg \textit{\url{vehicle42.vw.}}) and each service use unique keys, which are independently derived. 
The \ac{ZSK} per vehicle is signed by the \ac{OEM} \ac{KSK}, building the chain of trust as visualized in \cref{fig:dnssecsigning}.

Using unique key pairs per service in each vehicle presents a strongly enhanced security solution but poses scalability challenges.
For example, assuming 10 million cars manufactured per year over a decade, each with 500 services, the system would need to handle at least 50 billion entries. 
This problem is not unique to our \ac{DNSSEC} approach but affects any solution where an \ac{OEM} maintains all vehicle credentials.
Here, the distributed and delegated DNS provides a distinct advantage.
We deploy a dedicated \ac{DNSSEC} zone per car (see \cref{fig:dnssecsigning}) with only the required few thousand records making it easily manageable, compared to Internet \ac{DNSSEC} deployment.

\customparagraph{Differentiation from pre-deployed certificates.}
Our certificate management process enables dynamic service and certificate deployment, addressing the limitations of static, pre-deployed certificates. 
Pre-deployed certificates require all parties to know public certificates in advance and leads to cascading updates whenever services are modified. 
When a service is updated or added, it must be deployed not only with its own keys but also with public certificates for all other services that may interact with it.
Equally, interacting services must be updated with the new public certificates, which generates maintenance overhead.
In contrast, our approach dynamically requests certificates via \ac{DNSSEC} when new communication partners or services are discovered, eliminating the need for prior knowledge of the full in-vehicle service deployment.
Once obtained, these credentials function like pre-deployed certificates over their \ac{TTL}. 

\subsection{Online vs. Offline Operation}
\label{sec:offlineops}
Primarily designed for \textbf{intermittent online operation}, our approach facilitates certificate updates and dynamic service deployments.
In the automotive domain, connectivity is now a de facto standard, with online features, such as eCall, roadside interactions, or charging stations. 
Our solution supports such online integration, since cars can individually authenticate based on credentials in \ac{DNSSEC}. 

\customparagraph{Fallback offline operation.}
Vehicles may operate offline for extended periods, \eg during long-distance travel or in areas with limited connectivity.
While offline scenarios reduce the remote attack surface, malicious payloads could be pre-deployed and stealthily execute while the vehicle is offline.
The recursive resolver supports offline verification by caching and validating \ac{DNSSEC} records locally, eliminating the need for external infrastructure access during standard vehicle operation. 
\ac{DNSSEC} record updates are restricted to online periods when services or credentials are refreshed in the (resting) car. 

To ensure reliability and security during long offline periods (\ie weeks or months), record and certificate lifetimes must be sufficiently long. 
The vehicular use case may require values that are uncommon in the global Internet, nevertheless largely improves agility over pre-installed certificates. 
Session resumption via last-known keys can defer re-keying until a reconnect.
For prolonged offline operation, degradation modes can maintain basic functions, such as extended use of expired credentials or special fallback firmware loaded from a secure boot.
The design of such a fallbacks may approximate present firmwares of unconnected cars. 

\customparagraph{End of life considerations.}
When \acp{OEM} or third-party suppliers discontinue support for older vehicle generations, certificate expiration becomes a critical issue. 
Mitigation strategies include using self-signed certificates, establishing an in-car trust chain, or deploying fallback firmware with pre-installed certificates. 
While the degradation of functionality and potential decrease in operational lifetime of cars that go out of \ac{OEM} support is a concern, it is not specific to our proposed security solution.
Future (possibly autonomous) connected cars will demand security maintenance---an area where DNSSEC offers a long-living, well-established, and robust foundation.

\section{Security Analysis}
\label{sec:security}
We evaluate the security properties against common threats and known attacks on automotive publish-subscribe systems.
We rely on established cryptographic algorithms and protocols (\eg \ac{DH}) without introducing new crypto primitives. 
\ac{DNSSEC} protocols are backed by formal analysis~\cite{k-vddjr-14,ldhbv-ffejr-23,zskbl-pdcjr-25} and cater for state of the art cryptography with extensions to post-quantum~\cite{mjhor-rpqjr-20}.
Hence, we do not consider attacks on keys, primitives, or crypto algorithms and assume \ac{DNSSEC} secure.
Instead, we focus on security properties of the protocol design and its integration with the automotive middleware \ac{SOME/IP}.
We first identify key \ac{DNSSEC} properties that are critical for automotive security, then analyze protocol state transitions with STRIDE threats and identify open limitations.

\customparagraph{Adversary and threat model.}
\label{sec:threatmodel}
We analyze automotive cybersecurity threats~\cite{mbzls-ssajr-18,zprsk-tmajr-22,atm} and assume attackers with critical knowledge and advanced tools to exploit the entire vehicle attack surface~\cite{u-2gajr-24,cmkas-ceaas-11,mv-reupv-15,zprsk-tmajr-22,mbzls-ssajr-18,yl-sacjr-20}. 
Attackers can compromise any device and bypass connectivity gateways to gain remote access to all communication channels.
We apply the Dolev-Yao adversary model~\cite{dy-spkjr-81} for public key protocols, previously used for automotive networks~\cite{mpmsl-sanjr-17}. 
Here, the adversary can intercept, modify, replay, block, or inject messages on all communication channels. 
However, we assume that an attacker cannot access securely stored key material, or change physical network connections~\cite{ii-fadjr-24}. 
Additional measures such as secure flow control~\cite{hmks-stsnv-23}, intrusion detection~\cite{rmwh-sadjr-18}, and firewalls~\cite{psz-hcdjr-17} may be in place but are excluded from our evaluation.

\subsection{Threat Protection Inherited from DNSSEC} 
Our automotive integration preserves the security properties of DNSSEC, \ac{DANE}, and \ac{DANCE}.

\customparagraph{Certificate security}
\label{sec:security-tlsa-cert}
relies on the DANE TLSA record, which binds certificates to DNSSEC-protected names, ensuring integrity and authenticity~\cite{RFC-6698, RFC-7671}. 
By adopting \codeword{DANE-EE} certificate usage, we eliminate third-party \ac{CA} dependencies, as the TLSA record directly references the certificate. 
This lean verification avoids potential inconsistencies and ensures conflict-free validation~\cite{RFC-7673}, as far as the DNSSEC infrastructure remains secure.

\customparagraph{Record integrity and caching.}
\label{sec:security-cache}
\ac{DNSSEC} record provisioning and validation is a critical function in our system architecture and performed by the onboard recursive resolver. 
This DNS resolver is considered a trusted component of the car, on which services rely on start-up to ensure the availability and integrity of certificate records. 
Still, service endpoints retain the option to validate record integrity themselves, which could be part of a routine monitoring or triggered by an alert~\cite{lamrk-uhdms-22}.

DNS communication with the outside world uses a secure channel to prevent man-in-the-middle attacks. 
The onboard resolver will also require common protective measures (\eg firewall filters) to prevent remote access and misuse in DDoS attacks, plus rate limiting of internal queries to prevent overload or complexity attacks by malicious ECUs. 
As any certificate server, the DNSSEC resolver is a critical component, the protection of which is eased by mature implementations and decades of operational practice in the open Internet.

\customparagraph{Key management and infrastructure resilience.}
\label{sec:security-dnssec-keys}
DNSSEC relies on the security of DNSKEYs, which depends on technical safeguards, management processes, and personnel expertise~\cite{mtwhc-rrryr-19}. 
The separation of \acp{KSK} and \acp{ZSK} enhances security, as \acp{KSK} can be stored offline. 
The domain-constrained usage of DNSKEYs limits the cross-domain impact of attacks, as each key is restricted to its managed zone.
In our design, each vehicle operates in its individual zone, which prevents cross-vehicle escalations.
The private vehicular DNSSEC namespace removes reliance on public infrastructure and external trust anchors.
Still, a robust DNSSEC infrastructure requires automated and secure management processes~\cite{crccl-leejr-17,lgrkc-lcsde-20}, which are well established. 
The automotive \acp{OEM} in our design retain sole control of their DNS trees, which should enable a professional management. 

\label{sec:security-dnssec-infrastructure}

\begin{table*}
\centering
\caption{Summary of our STRIDE threat analysis of the proposed DNSSEC-based authentication and authorization scheme.} 
\label{tab:security_analysis}
\setlength{\tabcolsep}{3pt} 
\renewcommand{\arraystretch}{1.2} 
\small
\begin{tabularx}{\linewidth}{l L L}
\toprule
\textbf{STRIDE threat} & \textbf{Automotive attack vector} & \textbf{Protective measure} \\
\midrule
\textbf{S}poofing & Impersonation of publisher or subscriber identities, \eg conflicting offers or cancellation of subscriptions & Challenge response handshake; signatures validated against TLSA records \\
\textbf{T}ampering & Manipulation of discovery or subscription messages, \eg man-in-the-middle, replay attacks &Signing critical discovery messages; nonces and MACs as proposed in~\cite{irrsv-ssvjr-20,myzzc-ascjr-22}; encrypted data from derived session key \\
\textbf{R}epudiation & Denying involvement in a transaction or communication, \eg a subscription request or data exchange & Name-based authorization proves the origin of messages; logging actions prevents denial \\
\textbf{I}nformation disclosure & Unauthorized access to sensitive information, \eg service relationships and data exchange & Authorization along name hierarchy; DNSSEC protects against false certificate associations; secure channel \\
\textbf{D}enial of service & Disrupt service availability, \eg exhausting resources, or exploiting state transitions and timings in protocols & Spoofing mitigation; SOME/IP stack configurations; rate limiting; DNSSEC resolver protections \\
\textbf{E}levation of privilege & Gaining higher-level access or permissions than authorized, \eg access to safety critical domains & Scoped name hierarchy; TLSA association to specific names, protocols, ports; OEM controlled trust chain\\
\bottomrule
\end{tabularx}
\end{table*}

\subsection{STRIDE Threat Analysis}
\label{sec:stride}
We apply the STRIDE model~\cite{hl-sdljr-06} to analyze threats in an automotive \ac{SOA} that uses \ac{SOME/IP} as a case study.
STRIDE covers relevant attack vectors for distributed systems communication, such as \textbf{S}poofing, \textbf{T}ampering, \textbf{R}epudiation, \textbf{I}nformation disclosure, \textbf{D}enial of service, \textbf{E}levation of privilege.
We discuss how our mechanisms mitigate these risks, summarized in \cref{tab:security_analysis}.

\customparagraph{Spoofing}
\label{sec:security-spoofing}
of publisher or subscriber identities is a major concern during the service discovery by multicast~\cite{ii-fadjr-24,zlkk-assjr-21}.
We authenticate the previously unprotected SOME/IP \codeword{Offer}, \codeword{Subscribe}, and \codeword{SubscribeAck} messages by a challenge-response scheme.
We do not secure \codeword{Find} messages, since every endpoint can discover all services as per design of SOME/IP and DNS.
To prevent false service offers, subscribers verify the service endpoint using SVCB records and subscribe only to matching endpoints. 
Challenges included in offer and subscribe messages request mutual authentication of each party.
All message signatures are validated against (DNSSEC-protected) TLSA records, which ensures the authenticity and authorizes the publisher and subscriber, thereby preventing spoofing.

\customparagraph{Tampering}
\label{sec:security-tampering}
with discovery or subscription messages invalidates signatures, against which our scheme protects.
Following successful initial verification, an integrated \ac{DH} key exchange~\cite{dh-ndcjr-76} establishes session keys that ensure confidentiality and integrity for all subsequent communication.
Signed nonces mitigate simple replay attacks by enforcing single-use processing.
Signatures and \acp{MAC}~\cite{irrsv-ssvjr-20,myzzc-ascjr-22} ensure message integrity and thus protect against man-in-the-middle attacks. 
Endpoints in discovery messages are immutable, as they are cryptographically authenticated.
Cyclic subscription renewals and retries mitigate simple drop attacks, though targeted blackholing (\eg at the switch level) or hardware defects cannot be mitigated on the message authentication layer.

\customparagraph{Repudiation.}
\label{sec:security-repudiation}
All actor identities are authenticated and authorized prior to service access. 
In the presence of event logging, repudiation attacks can be identified and attributed. 
Name-based authorization supports differentiated service access, which shall prevent repudiation in the absence of compromised components. 

\customparagraph{Information disclosure.}
\label{sec:security-disclosure}
A service authorizes the name of the subscriber for access---names are protected via valid TLSA records. 
Data is only transmitted after mutual authentication and authorization of both parties and after an encrypted channel with perfect forward secrecy is established via the \ac{DH} key establishment. 
In multicast scenarios, publishers share the session key over this secure channel with valid subscribers.

The \ac{DNSSEC} itself lacks protective privacy like the DNS, which opens the potential of domain discovery and thereby the disclosure of active services in the car. 
In the presence of service discovery, this raises no concern within the car.
For external protection, the \acp{OEM} retain flexibility to shield both the (private) DNS infrastructure and local resolvers, balancing ease of implementation, performance, and confidentiality.

\customparagraph{Denial of service.}
\label{sec:security-dos}
Service availability is a safety requirement for driving vehicles. Current automotive systems have little redundancy but shall only operate after critical services are started and communication is established.
Prior, the service discovery phase is not time-critical and service availability is not guaranteed within the design of current cars. 
For the SOME/IP service discovery, \ac{DoS} attacks have been identified~\cite{ii-fadjr-24,zlkk-assjr-21}, mainly abusing state transitions or timings in the protocol by spoofing messages. 
Spoofing, however, is mitigated by our authentication scheme. 
Additional protective measures can be configured for the SOME/IP stack~\cite{ii-fadjr-24}.

\ac{DNSSEC} resolution may delay the service discovery. 
Mass requests at the local resolver may lead to \ac{DoS}, however, current resolvers are optimized for many parallel requests and protected by rate-limiting. Complexity attacks that target the DNSSEC resolver are possible, \eg requests for \textit{existing} records with long DNSSEC chains can cause a larger number of crypto operations. 
This can be prevented by restricting resolver access to the pre-fetched zone relevant for car operations. \acf{TSN} protocols~\cite{ieee8021q-22}, as foreseen in future cars, will additionally protect time-critical streams from delays by non-time-critical DNS requests.
The in-car resolver is only accessible from within the car; thus, external services cannot exploit the DNS for attacks (\eg reflective amplification).

\customparagraph{Elevation of privilege.}
\label{sec:security-elevation}
The publisher uses the DNS name of the subscriber to authorize the subscription during authentication.
TLSA certificate associations are bound to names, protocols, and ports, which limits the attack surface. 
DNSSEC-protected records prevent elevation of privileges, as the correct private key is required to impersonate a service.
By strict separation of concerns, the OEM can manage a private DNSSEC infrastructure, maintaining the chain of trust independently of service suppliers and Internet trust anchors.

\subsection{Discussions and Limitations}
Our approach successfully addresses authentication and authorization using the established Internet standards DNSSEC, DANE, and DANCE for robust credential lifecycle management. 
With this, we protect SOME/IP against the common STRIDE threats. 
Still, some limitations are to be discussed. 

\customparagraph{Mixed security levels.}
Our analysis focuses on providing the highest security level for automotive service discovery, which may be unneeded for some services. 
Different levels of security are likely~\cite{irrsv-ssvjr-20}, \eg not all services require data encryption. 
Allowing for mixed security statuses improves backward compatibility with existing systems. 
Subscribers and publishers might request authentication, authorization, or encryption depending on service requirements~\cite{irrsv-ssvjr-20}. 
Nevertheless, for mixed security levels, downgrade attacks, in which adversaries enforce a weaker security model, demand separate mitigation~\cite{irrsv-ssvjr-20,RFC-6698} and are not addressed here.

\customparagraph{Supply chain attacks.}
We decouple credential management (secured via \ac{OEM}-signed DNSSEC) from software distribution (supplier-signed).

Vehicles fully rely on trust in the \ac{OEM} via the DNSSEC chain without process attestation. 
Secure build systems, \ac{OTA} updates~\cite{mns-stajr-22}, and bootstrapping~\cite{sksb-sbrjr-20} can protect the automotive software supply chain, but must be independently asserted. 

\customparagraph{Private key compromise.}
We consider private keys safely protected by \acp{HSM}.
If a private publisher key gets compromised, the attacker could offer services authenticated by a bogus certificate. 
Nevertheless, our approach restricts fraudulent offers to services with valid endpoint information in the SVCB records and a matching TLSA record, which require the cooperation of the OEM. 
Similarly, a compromised private key of a subscriber enables the attacker to access services associated with TLSA records tied to the compromised client key. 
We emphasize that each vehicle uses unique keys for its service zone, so that a compromise only affects the specific vehicle deployment. 

\customparagraph{Discovery delays.}
While spoofing and most \ac{DoS} attacks are mitigated, timing attacks may still introduce discovery delays~\cite{ii-fadjr-24}. 
For example, upon receiving an offer message, a subscriber queries the DNSSEC resolver for the respective TLSA and SVCB records. 
During this resolution period, attackers could inject conflicting offers for the same publisher. 
Under the current \ac{SOME/IP} design (see \cref{fig:fsm_client_sd}), only the most recent offer undergoes validation once DNSSEC records are resolved, which will fail in this case. 
In the worst case, subscribers can validate the next valid cyclic offer, potentially delaying discovery by one full cycle. 
Similarly, publishers may receive multiple subscribe messages for the same client before the client certificate is obtained. 
Although caching of pending requests could reduce such delays, this approach may introduce new attack vectors. 
Ultimately, all conflicts resolve once DNSSEC records are retrieved, allowing for immediate authentication of requests, which limits the worst-case discovery delay. 
Initializing secure connections before vehicle operation further reduces the impact of these delays.

\customparagraph{Complexity attacks}
become a threat if an attacker can trigger or inflate cryptographic operations that overload the system. 
In the regular mode of the car, all asymmetric crypto operations of DNSSEC are asynchronously executed by the internal resolver. 
Update triggers only fire if DNS records change, the rate of which should be internally limited within a well configured DNSSEC infrastructure.
At the service endpoints, nonce generation, signature generation, and verification happen only in the initial phase of session establishment. 
Corresponding complexity attacks in this phase will delay or suspend the startup of the car. 
Vehicles could mitigate remote attacks by closing the gateway or tightening the firewall. 
In the case of an attacking internal component or service instantiation failures, the vehicle would likely require a software reset (rollback, either \ac{OTA} or via offline service maintenance.
Since authentication occurs pre-drive, safety-critical sessions---as well as such a reset---are established at start-up, ensuring real-time operation remains unaffected. 

State explosion attacks exhaust memory for service offers and subscriptions. 
While states for service offers and active subscriptions remain compatible to SOME/IP, our solution increases the memory footprint by additional nonces, certificates, and signatures. 
For unauthenticated subscription requests, DNSSEC validation requires to store requests until completion. 
An attack requires a high volume of malicious requests and can be mitigated by throttling request storms.

\begin{table*}
\centering
\caption{Publisher (service \textit{42}, instance \textit{1}, version \textit{2.3})
and subscriber (client \textit{17}) records in \textit{\url{vehicle1.oem.}} zone.
}
\small
\begin{tabularx}{\linewidth}{llL}
\toprule
\textbf{Name} & \textbf{Type} & \textbf{Value} \\
\midrule
\small{\url{_someip.3.2.1.42.service.vehicle1.oem.}} & \small{IN SVCB} & \small{1 . ipv4hint=10.0.0.2 port=5000 instance=1 major=2 minor=3 ip\_proto=17}\\
\small{\url{_5000._someip.3.2.1.42.service.vehicle1.oem.}} & \small{IN TLSA} & \small{3 0 0 (308205...)} \\
\small{\url{_someip-client.2.1.42.17.client.vehicle1.oem.}} & \small{IN TLSA} & \small{3 0 0 (308204c...)} \\
\bottomrule
\end{tabularx}
\label{tab:dns_sample_records}
\end{table*} 

\section{Evaluation in a Realistic In-Car Scenario}
\label{sec:evaluation}
We compare the impact of DNSSEC-based authentication and authorization against two baselines.
\one\textit{Vanilla vsomeip}, the reference implementation with additional statistics to measure performance.
\two\textit{Pre-deployed certificates}, a full publisher and subscriber authentication using static certificates.
\three\textit{DNSSEC, DANE, and DANCE}, our DNSSEC-based authentication \textit{and} authorization with local DNS resolution.
We focus on service discovery performance, cryptographic overhead, and DNSSEC resolution latency.

Our implementation builds upon open-source components, including the \textit{vsomeip}~\cite{vsomeip} reference implementation  (see \cref{sec:someip}).
Extensions for our authentication scheme rely on libraries for DNS lookups (\textit{c-ares}~\cite{c-ares}) and cryptography (\textit{Crypto++}~\cite{cryptopp}).
All services run on a VM (\textit{Ubuntu} 22.04.5 LTS, 32 GB RAM, 16 Intel\textcopyright 13900K p-cores). 
We employ \textit{Mininet}~\cite{mininet} as a virtual network and \textit{NSD}~\cite{nsd} as our DNSSEC resolver.
For service discovery, we use default \textit{vsomeip} settings:
an initial delay between \qtyrange{10}{100}{\milli\second},
cyclic offers every \SI{2}{\second},
and up to 3 find/offer retransmissions with delays doubling per attempt starting from \SI{200}{\milli\second}.
Our implementation, evaluation scenarios, and data processing scripts are available as open-source at \url{https://github.com/inetrg/dnssec-for-cars}.

\subsection{Configuring a Realistic IVN}
We use our open-source \ac{IVN} model~\cite{mhlks-fsaad-24} of a production vehicle converted to a zone topology. 
To simplify the Mininet implementation, we replaced the original ring backbone with a star topology (see \cref{fig:ivn_topology}).
The network consists of five switches, four lidar systems, two cameras, four zone control \acp{ECU}, and three \acp{HPC} for \ac{ADAS}, infotainment, etc.
All original communication relationships are preserved when converting applications to SOME/IP publishers and subscribers. 

The resulting configuration comprises 212 publishers and 448 subscribers
with an average of \SI{2.1}{} subscribers per publisher (min 1, max 4), exclusively using multicast endpoints. 
On the node level,  an average of 16 local publishers run per node (min 0, max 79) and serve 35 remote subscribers on average (min 0, max 174).
Simultaneously, each node hosts an average of 35 local subscribers (min 0, max 131), with 35 remote publishers (min 0, max 79).

For secure communication, all publishers and subscribers use unique key pairs generated with OpenSSL~\cite{openssl} (SHA-256, 2048-bit RSA keys). 
\textit{Vanilla vsomeip} does not support certificates.
\textit{Pre-deployed certificates} are statically linked in the application configuration. 
In the \textit{DNSSEC} scenario, publishers and subscribers store only their private keys and resolve certificates via DNSSEC.

We establish a dedicated DNSSEC zone for the vehicle under the domain \textit{\url{vehicle1.oem.}}. 
The complete \ac{IVN} requires 872 SVCB and TLSA records, each cryptographically signed with an RRSIG record.
DNSSEC clients rely on the resolver for validation.
\cref{tab:dns_sample_records} shows example records for one publisher with one subscriber.
Each of the 212 publishers has an SVCB record that describes the publisher endpoint within custom key-value pairs following RFC 9460~\cite{RFC-9460}. 
It encodes SOME/IP discovery metadata (\eg service identifiers, instance numbers, and version information) as we proposed in~\cite{mhmks-asasd-23}.
All TLSA records (448 subscribers, 212 publishers) are specific to the full service information, and we use the \codeword{DANE-EE} certificate association type, which directly authenticates service certificates without intermediate \acp{CA}.
We do not utilize wildcards, as we cannot realistically derive them from the original anonymized car network~\cite{mhlks-fsaad-24}. 

\subsection{Performance Evaluation}
We evaluate the performance in our scenario of 660 applications initialized and emulated on a single host system, which is challenging. 
\cref{tab:results} shows performance results collected across 25 independent runs for each scenario. 
Our statistical analysis extracts minimum, mean, and maximum values to span variability across our 16-core system.

\begin{table*}
    \centering
    \caption{Startup latencies across the three scenarios (in \si{\milli\second}).}
    \label{tab:results}
    \small
    \begin{tabularx}{\linewidth}{L ccc ccc ccc}
    \toprule
    \textbf{Metric} 
        & \multicolumn{3}{c}{\textbf{Vanilla vsomeip}}
        & \multicolumn{3}{c}{\textbf{Pre-deployed}}
        & \multicolumn{3}{c}{\textbf{DNSSEC}}
    \\ \cmidrule(r){2-4} \cmidrule(lr){5-7} \cmidrule(l){8-10}
    & \textbf{Min} & \textbf{Mean} & \textbf{Max} & \textbf{Min} & \textbf{Mean} & \textbf{Max} & \textbf{Min} & \textbf{Mean} & \textbf{Max} \\
    \midrule
    Full network startup & \SI{802.4}{} & \SI{1054.3}{} & \SI{1244.4}{} & \SI{1171.4}{} & \SI{1237.0}{} & \SI{1348.7}{} & \SI{1165.4}{} & \SI{1234.4}{} & \SI{1298.4}{}\\
    Total initialization time & \SI{780.0}{} & \SI{808.6}{} & \SI{845.8}{} & \SI{833.4}{} & \SI{852.6}{} & \SI{877.1}{} & \SI{848.2}{} & \SI{865.6}{} & \SI{883.6}{}\\
    Subscription setup & \SI{25.5}{} & \SI{70.6}{} & \SI{399.3}{} & \SI{129.9}{} & \SI{218.5}{} & \SI{670.2}{} & \SI{123.5}{} & \SI{226.2}{} & \SI{455.6}{}\\
    Crypto operations per subscription & N/A & N/A & N/A & \SI{1.2}{} & \SI{1.7}{} & \SI{46.9}{} & \SI{1.2}{} & \SI{1.8}{} & \SI{50.0}{}\\
    DNS resolution per subscription & N/A & N/A & N/A & N/A & N/A & N/A & \SI{0.3}{} & \SI{18.0}{} & \SI{168.4}{}\\
    \bottomrule
\end{tabularx}

\end{table*}

\emph{Full network startup} measures the time between the first recorded offer and the last recorded acknowledgment of the run.
It spans \SIrange{0.8}{1.3}{\second} across all scenarios.
Launching the environment and all applications (\emph{total initialization time}) attains relatively stable values with a $\approx 7$\% overhead for initializing crypto.
Communication begins directly after each application is initialized with random startup order and initial wait phases (\SIrange{10}{100}{\milli\second}). 
These random factors appear to dominate the cryptographic overhead.

The \emph{subscription setup} is measured from the initial publisher offer to the acknowledgment received by the subscriber, including message passing, all cryptographic operations, and DNSSEC lookups.
Startup order and wait phases significantly influence this metric, since publisher-subscriber pairs may launch with delay and miss initial offer messages, which requires them to wait for the next offer cycle or find request. 
The observed variability appears counterintuitive, since the \textit{pre-deployed certificate} does not outperform the \textit{DNSSEC, DANE, and DANCE} approach---likely due to system fluctuations.

Both crypto-enhanced scenarios include nonce generation, message signing/verification, and hash generation (summed as \emph{crypto operations per subscription}). 
As expected, the setup time per subscription increases with the introduction of our security scheme.
The performance of crypto operations remains comparable between the two scenarios, showing negligible overhead of additional validation steps in the DNSSEC variant. 
While a mean cryptographic overhead of \SI{1.7}{\milli\second} is low relative to the \SIrange{10}{100}{\milli\second} initial configuration delay, our multi-service \acp{ECU} experience subscription floods on every offer cycle. 
This can lead to large outliers, likely caused by the execution of many crypto operations sequentially on the same node and in parallel on different nodes in our one machine setup.
Multi-service \acp{HPC} systems may require optimizations, such as parallel execution of crypto operations or \ac{HSM} support for efficiency.

The DNSSEC-based authentication and authorization scheme requires two lookups per publisher and one per subscriber.
These lookups take a minimum of \SI{0.3}{\milli\second}, which is expected for local resolution on the same machine.
Higher mean and maximum times likely result from query processing and queueing on the nodes and the resolver.
Limitations on the number of channels for parallel DNSSEC resolution may cause delays at peak times. 
Overall, we diagnose that the maximum startup delay inherited from additional overhead of the DNSSEC scenario remains within a few 100 ms, which is hardly notable by the driver of a car. 

\subsection{Discussions and Limitations}
The evaluation demonstrates the feasibility of DNSSEC-based authentication and authorization in a realistic \ac{IVN}. 
The results provide estimates for real-world deployment in an \ac{OEM} fleet, in which each vehicle zone requires approximately \SI{1000}{} records. 
Unlike alternative solutions, private DNSSEC zones enable secure isolation of record and certificate management within a scalable, \ac{OEM} controlled infrastructure that prevents certificate association conflicts.

\customparagraph{Limitations of the evaluation setup.}
Our evaluation aims to accurately replicate a realistic vehicle in an emulation environment. 
While our emulated Open vSwitch in Mininet does not account for (insignificant) network transmission times, 
the virtualized system is similar to the virtualized architectures of \ac{HPC}-centered computing environments in new vehicle generations~\cite{a-naejr-24,b-fsnjr-25}.
However, our configuration uses a single host of desktop grade, which may differ from automotive hardware. 

The evaluation setup showed performance jitter primarily attributable to random fluctuations while executing 660 applications on a shared 16-core CPU, dominated by concurrent initialization. 
Benchmarks of smaller configurations showed that individual service setup times are not strongly affected by subscriber count and can remain well below the \SI{200}{\milli\second} requirement~\cite{hmks-stsnv-23}.
Future investigations in a truly distributed setup with real automotive hardware and communication traces could provide further insights.

Our evaluation showed that the DNSSEC-based approach adds little to no overhead to pre-deployed certificates. 
As service discovery is not safety critical, an average increase in network startup time by \SI{17}{\percent} for cryptography is acceptable.
Optimizations of scattering the discovery messages, resolving the DNS, or parallelization could circumvent crypto bottlenecks.
Most of these challenges are generic and also affect the \textit{vanilla vsomeip} and \textit{pre-deployed certificates} baselines.

\customparagraph{DNSSEC resolution}
times for single requests are very small in a local environment.
DNSSEC resolvers are well-optimized for resolving concurrent queries, which limits the delay of a single query.
Nevertheless, our implementation is not optimized for concurrent operations of mass requests, which produced significant outliers. 
Our evaluation excluded remote DNSSEC cache population, the performance of which is well known.
We anticipate remote DNSSEC resolution to be scheduled during rest or update periods rather than during the critical startup---without operational impact.

\customparagraph{Credential management.} 
Our car evaluation network indicates that publisher and subscriber nodes resolve 70 TLSA certificates on average via DNSSEC. 
With the \ac{SaaS} transformation of the automotive industry, we expect this number to grow as more services become volatile, regularly updated, or installed post-sale. 
From this perspective, it becomes apparent that the traditional pre-deployment of certificates faces severe management challenges.  
Nodes must securely store credentials even for applications not currently installed, to ensure compatibility with potential future updates or newly added services. 
This also introduces security risks, as discussed in \cref{sec:cert_management}.

Reducing the number of certificates (\eg to one per ECU) could ease deployment but would significantly limit service mobility and reduce the granularity of access control.
Our approach singles out certificate management to secure maintenance in the DNSSEC zone, which enables on-demand access to required credentials.
For performance, nodes can also cache DNS responses to become comparable with \textit{pre-deployed certificates}, while honoring the \ac{TTL} of records.

\customparagraph{Session resumption.}
In many cases \acp{IVN} remain unchanged between consecutive trips, which makes full cryptographic handshake at every startup redundant. 
Session resumption, \ie the reuse of previous session keys, could then eliminate the overhead of cryptographic handshakes. 
Performance of periodic offer and subscribe messages could thereby approximate \textit{vanilla vsomeip} without compromising security. 
Session reinitializations could be performed at rest periods.

\customparagraph{Safety considerations with TSN.}
For simplicity, our network evaluation did not include \ac{TSN} \ac{QoS} mechanisms, such as traffic shaping for time critical flows.
Our solution does not add delays to car operations and subsequent real-time communication, since authentication and session establishment occur only during startup---all services and sessions are in place for driving. 
Hence, deployment of TSN remains unchanged and aligns with results of prior work~\cite{smmks-nslld-26} about interactions between automotive \ac{SOA} and \ac{TSN}.
Overall, \ac{DNSSEC}-based security supplies credentials without degrading \ac{TSN} behavior.


\section{Conclusions and Outlook}
\label{sec:conclusion}

Future connected cars urgently need a robust and agile credential management system for authentication and authorization of their services. 
In this paper, we designed, implemented, and evaluated a key management framework for automotive publish-subscribe systems that provisions credentials via DNSSEC.
The framework decouples service provider keys from service deployment keys using DANE TLSA records. 
Each vehicle uses unique key pairs, authenticated against its own DNSSEC zone. 
Our security and performance evaluations revealed effective protection against common threats and scalable compliance with performance requirements of the onboard software system.

Our approach opens several future research directions. 
Performance evaluations in real-world cars will inspire protocol optimizations, including the exploration of post-quantum cryptography.
Session resumption techniques should further reduce overhead during vehicle startup.
The proposed security scheme can also be extend to a broader deployment including external roadside infrastructure (\ac{V2X}), charging stations, or a fleet management in the cloud.

\label{lastpage}

\appendix[Integration with Automotive SOA Protocols --- an Example of AUTOSAR SOME/IP]
\label{sec:someip}
Our security framework is designed to be compatible with a wide range of existing and emerging communication protocols by leveraging open standards.
We demonstrate this through an integration of DNSSEC-based authentication and authorization with \ac{SOME/IP}~\cite{a-spsjr-24,a-ssdjr-24}. 

\subsection{SOME/IP Messaging and Our Security Extensions}
\label{sec:someipheaders}

\begin{figure}
    \centering
    \includegraphics[width=\linewidth, trim=18pt 18pt 18pt 18pt, clip=true]{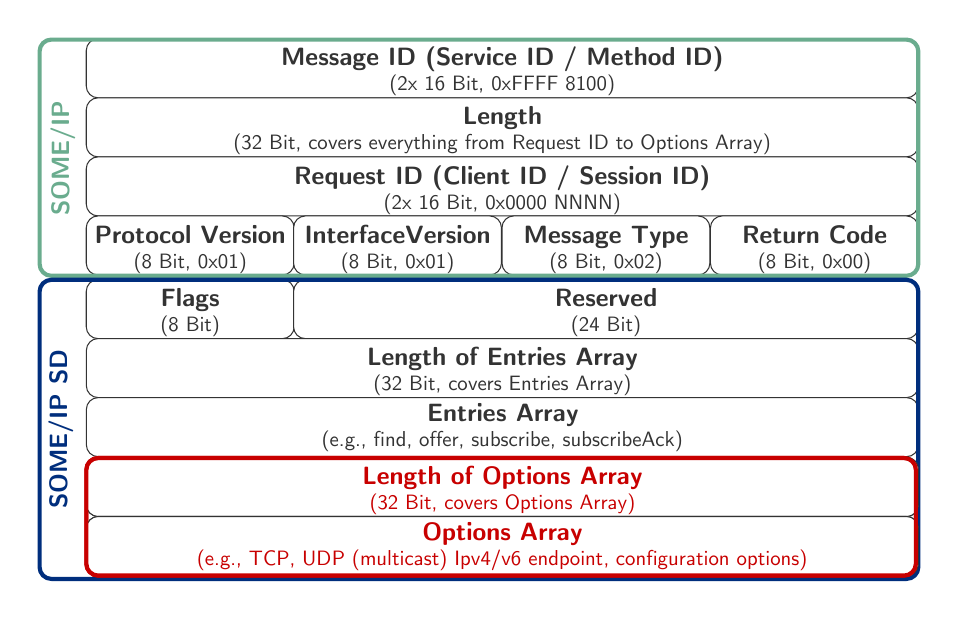}
    \caption{SOME/IP message layout for service discovery messages~\cite{a-ssdjr-24}.}
    \label{fig:someip_structure}
\end{figure}

\begin{figure}
    \centering
    \includegraphics[width=\linewidth, trim=18pt 18pt 18pt 18pt, clip=true]{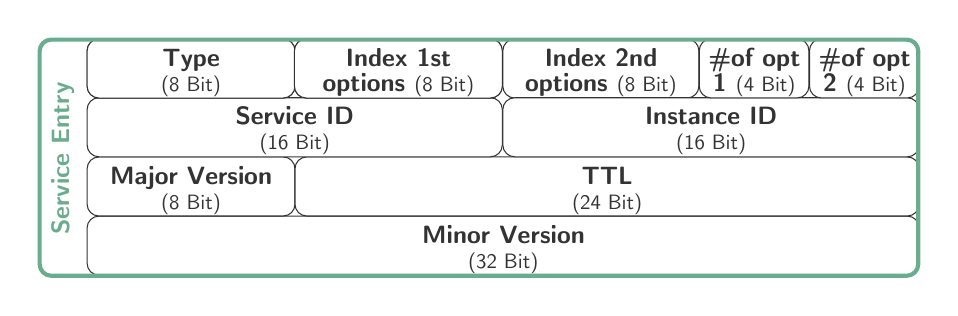}
    \caption{Structure of SOME/IP service discovery entry included in service discovery messages~\cite{a-ssdjr-24}.}
    \label{fig:someip_entry_structure}
\end{figure}

\ac{SOME/IP} employs a publish-subscribe model, using a find-offer mechanism to discover services and establish subscriptions.
Service discovery operates as a SOME/IP service that uses a known UDP-IP endpoint, \textit{Service ID}, \textit{Method ID}, and \textit{Client ID}. 
\cref{fig:someip_structure} depicts the message layout of discovery messages. 
They contain a list of entries (\codeword{Find}, \codeword{Offer}, \codeword{Subscribe}, \codeword{SubscribeAck}) and reference options that, for example, provide endpoint information.
\cref{fig:someip_entry_structure} shows the structure of entries.
\codeword{Offer} and \codeword{Subscribe} entries with a \ac{TTL} of zero indicate \codeword{StopOffer} and \codeword{StopSubscribe}.

\begin{figure}
    \centering
    \includegraphics[width=\linewidth, trim=10pt 18pt 18pt 18pt, clip=true]{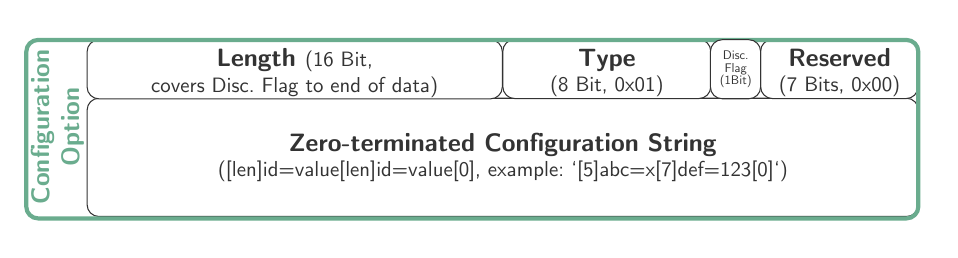}
    \caption{SOME/IP configuration option layout included in service discovery messages~\cite{a-ssdjr-24}.}
    \label{fig:someip_option}
\end{figure}

The standard supports custom \codeword{Configuration Options} structured as application-specific strings, depicted in \cref{fig:someip_option}.
We introduce four custom options:
\one a \textit{challenge} with a random nonce, \two a \textit{response} returning the signed nonce, \three \textit{key exchange} with the public \ac{DH} share and cryptographic parameters, and \four \textit{session key} with an encrypted symmetric key for confidential publishing and secure multicast communication.
Nonce lengths are configurable, allowing \acp{OEM} to balance security and computational overhead---\SIrange{16}{32}{\byte} are usually sufficient (\cf OAuth 2.0 and TLS 1.3).
We use \SI{16}{\byte} nonces for increased compatibility (timestamp: \SI{8}{\byte}; random value: \SI{8}{\byte}), which enforces limited validity and mitigates brute-force attacks.

\subsection{SOME/IP Service Discovery State Machines with Integrations for DNSSEC}
\label{sec:someipfsms}
\begin{figure}
    \centering
    \includegraphics[width=1.0\linewidth, trim=30pt 650pt 30pt 30pt, clip=true]{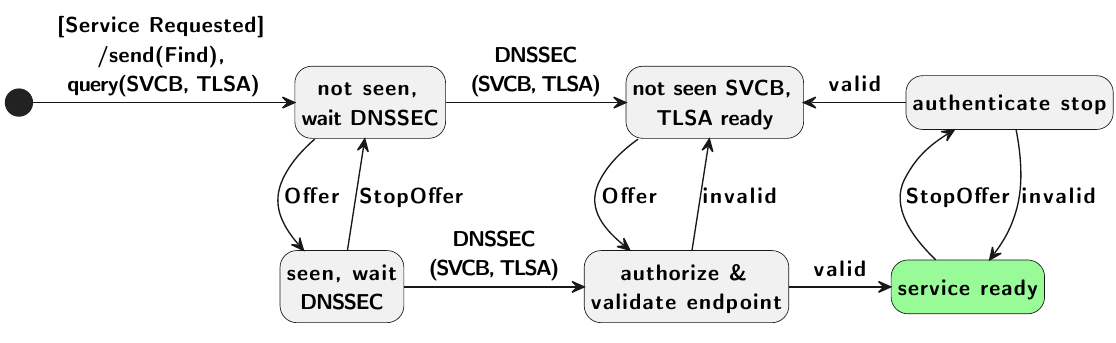}
    \caption{State machine for SOME/IP discovery of requested services at the client side including DNS interactions not including the initial wait phases, retransmissions, and timeout handling.}
    \label{fig:fsm_client_sd}
\end{figure}
\begin{figure}
    \centering
    \includegraphics[width=1.0\linewidth, trim=30pt 610pt 30pt 30pt, clip=true]{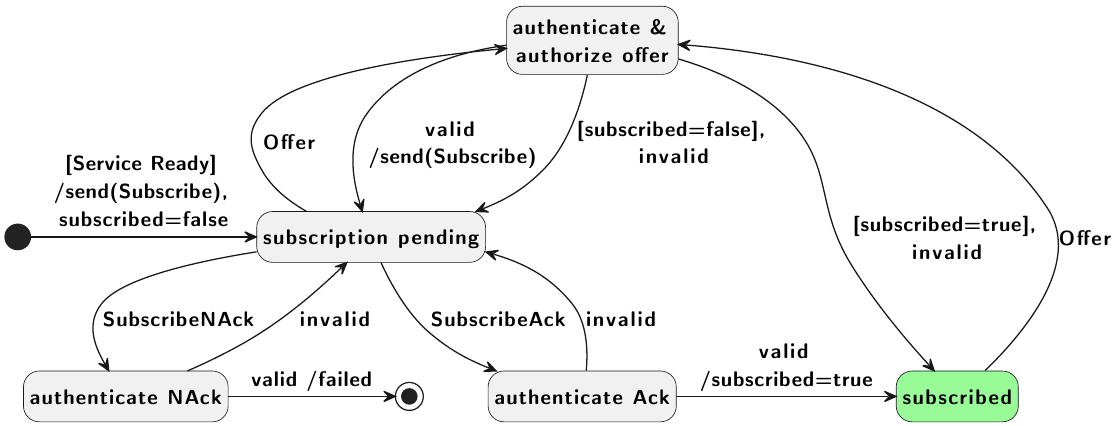}
    \caption{State machine for SOME/IP subscription handling with authentication and authorization at the client side not including, retransmissions, and timeout handling.}
    \label{fig:fsm_client_sub}
\end{figure}
\begin{figure}
    \centering
    \includegraphics[width=1.0\linewidth, trim=30pt 560pt 30pt 30pt, clip=true]{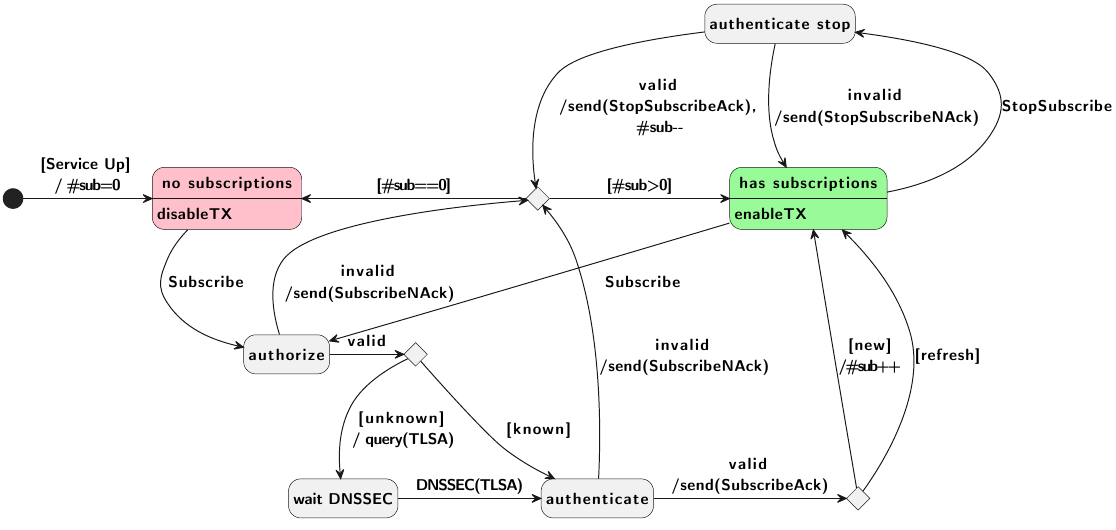}
    \caption{State machine for SOME/IP subscription handling at the server side including DNS interactions, authentication, and authorization, but not including, retransmissions, and timeout handling.}
    \label{fig:fsm_server_sub}
\end{figure}
The original SOME/IP protocol state machines are detailed in the specification~\cite{a-ssdjr-24}.
Our adaptations incorporate the proposed DNS interactions, and  the authentication and authorization mechanisms.
Figures \ref{fig:fsm_client_sd}, \ref{fig:fsm_client_sub}, and \ref{fig:fsm_server_sub} show the client-side discovery, subscription handling, and server-side subscription, respectively.
For clarity, we omit the initial wait phases, retransmissions, and timeout handling. 
We concentrate on scenarios where applications actively request services.
Practical details, such as hardware-specific events like interface status changes, are also not described.

Servers periodically send \codeword{Offer} messages upon startup, a behavior unchanged by the proposed security mechanisms.
Client-side discovery (see \cref{fig:fsm_client_sd}) starts with a \codeword{Find} message and a \ac{DNSSEC} query, with clients awaiting completion of both.
If DNSSEC lookups fail, the client may stop or proceed with an insecure subscription. 
Upon receiving an \codeword{Offer}, the client validates it against the SVCB record. 
Invalid \codeword{Offer}s are ignored, and the client awaits the next. 
\codeword{StopOffer} messages are also validated using the DNSSEC response.

Clients respond to (cyclic) \codeword{Offer}s by (re)issuing \codeword{Subscribe} messages.
This initiates subscription handling on the client-side (see \cref{fig:fsm_client_sub}) and clients await a \codeword{SubscribeAck}. 
At the server (see \cref{fig:fsm_server_sub}), incoming \codeword{Subscribe}s are first authorized by name. 
The server awaits the client TLSA record to authenticate the subscription and acknowledges it with a \codeword{SubscribeAck}.
Valid and active subscriptions are cached. 
At the client side, the acknowledgment is authenticated against the previously retrieved TLSA record. 
A valid acknowledgment establishes the subscription; otherwise, the client awaits another.

\bibliographystyle{IEEEtran}
\bibliography{bbl/ieee_style,bbl/bibliography,bib/own,bib/ids,bib/rfcs}

\end{document}